\documentclass[10pt,pdftex]{article}

\usepackage[english]{babel}
\usepackage{amsmath}
\usepackage{amssymb}
\usepackage[final]{pdfpages}
\usepackage{graphicx}
\usepackage[a4paper, text={16.26cm,23cm},centering]{geometry}
\usepackage{mathtools}

\usepackage[T1]{fontenc}
\usepackage[utf8]{inputenc}

\makeatletter

\normalsize
\renewcommand\section{\@startsection{section}{1}{\z@}%
                                   {-3.5ex \@plus -1.3ex \@minus -.7ex}%
                                   {2.3ex \@plus.4ex \@minus .4ex}%
                                   {\normalfont\large\bfseries}}
\renewcommand\subsection{\@startsection{subsection}{2}{\z@}%
                                   {-2.3ex\@plus -1ex \@minus -.5ex}%
                                   {1.2ex \@plus .3ex \@minus .3ex}%
                                   {\normalfont\normalsize\bfseries}}
\renewcommand\subsubsection{\@startsection{subsubsection}{3}{\z@}%
                                   {-2.3ex\@plus -1ex \@minus -.5ex}%
                                   {1ex \@plus .2ex \@minus .2ex}%
                                   {\normalfont\normalsize\bfseries}}
\renewcommand\paragraph{\@startsection{paragraph}{4}{\z@}%
                                   {1.75ex \@plus1ex \@minus.2ex}%
                                   {-1em}%
                                   {\normalfont\normalsize\bfseries}}
\renewcommand\subparagraph{\@startsection{subparagraph}{5}{\parindent}%
                                   {1.75ex \@plus1ex \@minus .2ex}%
                                   {-1em}%
                                   {\normalfont\normalsize\bfseries}}

\makeatother

\RequirePackage[sort&compress]{natbib}

\usepackage[pdftitle={Structure Formation in the Early Universe},
pdfauthor={P. G. Miedema}, pdfsubject={Structure Formation, Diabatic
  Perturbations}, pdfstartview={FitH}, pdfkeywords={Structure
  Formation, Cosmological Density Perturbations, Diabatic
  Perturbations, non-barotropic equations of state}, bookmarks,
bookmarksnumbered, colorlinks, linkcolor={blue}, urlcolor={blue},
citecolor={blue}]{hyperref}

\newcommand{\nul}{{\scriptscriptstyle(0)}}
\newcommand{\een}{{\scriptscriptstyle(1)}}
\newcommand{\true}{\text{phys}}
\newcommand{\aap}{A\&A} 

\begin{document}

\title{\textbf{\textsf{Structure Formation in the Early Universe}}}

\author{P.\ G.\
  Miedema\footnote{\href{mailto:PG.Miedema@ProtonMail.com}{PG.Miedema@ProtonMail.com}}
  \hfill\\
  Netherlands Defence Academy \\
  Hogeschoollaan 2 \\
  NL-4818CR Breda \\
  The Netherlands}


\date{March 9, 2021}

\maketitle

\begin{abstract}
  The evolution of the perturbations in the energy
  density and the particle number density in a flat Friedmann-Lema\^\i
  tre-Robertson-Walker universe in the radiation-dominated era and in
  the epoch after decoupling of matter and radiation is studied.  For
  large-scale perturbations the outcome is in accordance with
  treatments in the literature.  For small-scale perturbations the
  differences are conspicuous.  Firstly, in the radiation-dominated
  era small-scale perturbations grew proportional to the square root
  of time.  Secondly, perturbations in the Cold Dark Matter particle
  number density were, due to gravitation, coupled to perturbations in
  the total energy density.  This implies that structure formation
  could have begun successfully only after decoupling of matter and
  radiation.  Finally, after decoupling density perturbations evolved
  diabatically, i.e., they exchanged heat with their environment.
  This heat exchange may have enhanced the growth rate of their mass
  sufficiently to explain structure formation in the early universe, a
  phenomenon which cannot be understood from adiabatic density
  perturbations.
\end{abstract}


\begin{quote}
\textsc{pacs}: 98.62.Ai; 98.80.-k; 97.10Bt

\textsc{keywords}: Cosmology; structure formation; perturbation
theory; diabatic density perturbations
\end{quote}

\newpage
\hrule
\tableofcontents
\bigskip
\hrule

\section{Introduction}
\label{sec:intro}

The global properties of our universe are very well described by a
$\Lambda\textsc{cdm}$ model with a flat Friedmann-Lema\^\i
tre-Robertson-Walker (\textsc{flrw}) metric within the context of the
General Theory of Relativity.  To explain structure formation after
decoupling of matter and radiation in this model, one has to assume
that before decoupling Cold Dark Matter (\textsc{cdm}) has already
contracted to form seeds into which the baryons, i.e., ordinary
matter could fall after decoupling.  In this article it will be shown
that \textsc{cdm} did not contract faster than baryons before
decoupling and that structure formation started off successfully only
after decoupling.

In the companion article \cite{miedema:2014ada}\footnote{Section and
  equation numbers with a $\ast$ refer to sections and equations in
  the companion article.} expressions for the true, physical energy
density and particle number density perturbations
$\varepsilon_\een^\true$ and $n_\een^\true$ have been found.
Evolution equations for the corresponding contrast functions
$\delta_\varepsilon$ and $\delta_n$ have been derived for closed, flat
and open \textsc{flrw} universes.  In this article these evolution
equations will be applied to a flat \textsc{flrw} universe in its
three main phases, namely the radiation-dominated era, the plasma era,
and the epoch after decoupling of matter and radiation.  In the
derivation of the evolution equations, an equation of state for the
pressure of the form $p=p(n,\varepsilon)$ has been taken into account,
as is required by thermodynamics.  As a consequence, in addition to a
second-order evolution equation~(\ref{sec-ord}) for density
perturbations, a first-order evolution equation~(\ref{fir-ord}) for
entropy perturbations follows also from the perturbed Einstein
equations.  The coefficients of the second-order evolution equation
depend on the equation of state for the pressure and differs from
equations found in the literature. The entropy evolution equation is
absent in former treatments of the subject.  Therefore, the
system~(\ref{subeq:final}) leads to further reaching conclusions than
is possible from treatments in the literature.

Analytical expressions for the fluctuations in the energy density
$\delta_\varepsilon$ and the particle number density $\delta_n$ in the
radiation-dominated era and the epoch after decoupling will be
determined.  It is shown that the evolution
equations~(\ref{subeq:final}) corroborate the standard perturbation
theory in both eras in the limiting case of \emph{infinite}-scale
perturbations.  For \emph{finite} scales, however, the differences are
conspicuous.  Therefore, only finite-scale perturbations are
considered in detail.

In the appendix it is shown in detail why the standard perturbation
equation is inadequate to study the evolution of cosmological density
perturbations.  Furthermore, it is shown that the Jeans theory applied
to an expanding universe, i.e., the Newtonian cosmological
perturbation theory, is invalid.

\section{Einstein Equations and Conservation Laws for a Flat FLRW
  Universe}

In this section the equations needed for the study of the evolution of
density perturbations in the early universe are written down for an
equation of state for the pressure, $p=p(n,\varepsilon)$.

\subsection{Background Equations}

The set of zeroth-order Einstein equations and conservation laws for a
flat, i.e., $R_\nul=0$, \textsc{flrw} universe filled with a perfect
fluid with energy-momentum tensor
\begin{equation}
  \label{eq:en-mom-tensor}
  T^{\mu\nu}=(\varepsilon+p)u^\mu u^\nu - p g^{\mu\nu}, \quad 
p=p(n,\varepsilon),
\end{equation}
is given by~(8$^\ast$)
\begin{subequations}
\label{subeq:einstein-flrw}
\begin{alignat}{2}
  3H^2 & =\kappa\varepsilon_\nul, 
            & \kappa & =8\pi G_{\text{N}}/c^4,  \label{FRW3}\\
  \dot{\varepsilon}_\nul & = -3H\varepsilon_\nul(1+w),  
         \quad & w & \coloneqq p_\nul/\varepsilon_\nul,  \label{FRW2} \\
  \dot{n}_\nul & = -3Hn_\nul. && \label{FRW2a}
\end{alignat}
\end{subequations}
The evolution of density perturbations took place in the early
universe shortly after decoupling, when
$\Lambda \ll\kappa\varepsilon_\nul$. Therefore, the cosmological
constant $\Lambda$ has been neglected. An over-dot denotes
differentiation with respect to $ct$.

\subsection{Evolution Equations for Density Perturbations}

The complete set of perturbation equations for the two independent
density contrast functions $\delta_n$ and $\delta_\varepsilon$ is
given by~(65$^\ast$)
\begin{subequations}
\label{subeq:final}
\begin{align}
   \ddot{\delta}_\varepsilon + b_1 \dot{\delta}_\varepsilon +
      b_2 \delta_\varepsilon &=
      b_3 \left[\delta_n - \dfrac{\delta_\varepsilon}{1+w}\right],
              \label{sec-ord}  \\
   \dfrac{1}{c}\dfrac{{\text{d}}}{{\text{d}} t}
      \left[\delta_n - \dfrac{\delta_\varepsilon}{1 + w}\right] & =
     \dfrac{3Hn_\nul p_n}{\varepsilon_\nul(1 + w)}
     \left[\delta_n - \dfrac{\delta_\varepsilon}{1 + w}\right],
                 \label{fir-ord}
\end{align}
\end{subequations}
where the coefficients $b_1$, $b_2$ and~$b_3$,~(66$^\ast$), are for a
flat \textsc{flrw} universe filled with a perfect fluid described by
an equation of state for the pressure $p=p(n,\varepsilon)$ given by
\begin{subequations}
\label{subeq:coeff-contrast}
 \begin{align}
  b_1  & =  H(1-3w-3\beta^2)-2\dfrac{\dot{\beta}}{\beta}, \\
  b_2 & = \kappa\varepsilon_\nul\Bigl[2\beta^2(2+3w)-\tfrac{1}{6}(1+18w+9w^2) \Bigr]
      +2H\dfrac{\dot{\beta}}{\beta}(1+3w)-\beta^2\dfrac{\nabla^2}{a^2}, \\
  b_3 & = \Biggl\{\dfrac{-2}{1+w}
  \Biggl[\varepsilon_\nul p_{\varepsilon n}(1+w)
   +\dfrac{2p_n}{3H}\dfrac{\dot{\beta}}{\beta}
     +p_n(p_\varepsilon-\beta^2)+n_\nul p_{nn}\Biggr]+
   p_n\Biggr\}\dfrac{n_\nul}{\varepsilon_\nul}\dfrac{\nabla^2}{a^2},
\label{eq:b3}
\end{align}
\end{subequations}
where the partial derivatives of the equation of state for the pressure
$p(n,\varepsilon)$ are given by
\begin{equation}
  \label{eq:part-p}
  p_n\coloneqq\left(\dfrac{\partial p}{\partial n}\right)_{\!\!\varepsilon}, \quad
  p_\varepsilon\coloneqq\left(\dfrac{\partial p}{\partial \varepsilon}\right)_{\!\!n}, \quad
  p_{\varepsilon n}\coloneqq\dfrac{\partial^2p}{\partial\varepsilon\,\partial n}, \quad
  p_{nn}\coloneqq\dfrac{\partial^2p}{\partial n^2}.
\end{equation}
The symbol $\nabla^2$ denotes the Laplace operator.  The quantity
$\beta(t)$ is defined by
$\beta^2\coloneqq\dot{p}_\nul/\dot{\varepsilon}_\nul$.  Using that
$\dot{p}_\nul=p_n\dot{n}_\nul+p_\varepsilon\dot{\varepsilon}_\nul$ and
the conservation laws~(\ref{FRW2}) and~(\ref{FRW2a}) one gets
\begin{equation}
  \label{eq:beta-matter}
  \beta^2=p_\varepsilon+\dfrac{n_\nul p_n}{\varepsilon_\nul(1+w)}.
\end{equation}
From the definitions $w\coloneqq p_\nul/\varepsilon_\nul$ and
$\beta^2\coloneqq\dot{p}_\nul/\dot{\varepsilon}_\nul$ and the energy
conservation law~(\ref{FRW2}), one finds for the time-derivative of~$w$
\begin{equation}
   \dot{w}=3H(1+w)(w-\beta^2).  \label{eq:time-w}
\end{equation}
This expression holds true \emph{independent} of the precise form of
the equation of state.

The pressure perturbation is given by~(71$^\ast$)
\begin{equation}
    \label{eq:p-dia-adia}
  p_\een^\true=\beta^2\varepsilon_\nul\delta_\varepsilon+n_\nul
  p_n\left[\delta_n-\dfrac{\delta_\varepsilon}{1+w}  \right],
\end{equation}
where the first term, $\beta^2\varepsilon_\nul\delta_\varepsilon$, is
the adiabatic part and the second term the diabatic part of the
pressure perturbation.

The combined First and Second Law of Thermodynamics reads~(79$^\ast$)
\begin{equation}
  \label{eq:thermo-een}
  T_\nul s^\true_\een=-\dfrac{\varepsilon_\nul(1+w)}{n_\nul}
   \left[\delta_n-\dfrac{\delta_\varepsilon}{1+w}\right].
\end{equation}
Density perturbations evolve adiabatically if and only if the source
term of the evolution equation~(\ref{sec-ord}) vanishes, so that this
equation is homogeneous and describes, therefore, a closed system that
does not exchange heat with its environment.  This can only be
achieved for $p_n\approx0$, or, equivalently,
$p\approx p(\varepsilon)$, i.e., if the particle number density does
not contribute to the pressure.  In this case, the coefficient
$b_3$,~(\ref{eq:b3}), vanishes.

In view of~(\ref{eq:p-dia-adia}) and~(\ref{eq:thermo-een}), the
right-hand side of equation~(\ref{sec-ord}) will be referred to as the
entropy term and equation~(\ref{fir-ord}) is considered as the entropy
evolution equation.

\section{Analytical Solutions}
\label{sec:analytic-sol}

In this section analytical solutions of equations~(\ref{subeq:final})
are derived for a flat \textsc{flrw} universe with a vanishing
cosmological constant in its radiation-dominated phase and in the era
after decoupling of matter and radiation.  It is shown that $p_n\le0$
throughout the history of the universe.  In this case, the entropy
evolution equation~(\ref{fir-ord}) implies that fluctuations in the
particle number density, $\delta_n$, are coupled to fluctuations in
the total energy density, $\delta_\varepsilon$, through gravitation,
irrespective of the nature of the particles.  In particular, this
holds true for perturbations in \textsc{cdm}.  Consequently,
\textsc{cdm} fluctuations have evolved in the same way as
perturbations in ordinary matter.  This may rule out \textsc{cdm} as a
means to facilitate the formation of structure in the universe after
decoupling.  The same conclusion has also been reached by
\citet{2009arXiv0906.5087N}, on different grounds.  Therefore,
structure formation could start only \emph{after} decoupling.

\subsection{Radiation-dominated Era}
\label{sec:rad-dom-era}

At very high temperatures, radiation and ordinary matter are in
thermal equilibrium, coupled via Thomson scattering with the photons
dominating over the nucleons ($n_\gamma / n_\text{p}\approx
10^9$). Therefore the primordial fluid can be treated as
radiation-dominated with equations of state
\begin{equation}
  \label{eq:state-rad}
 \varepsilon=a_{\text{B}}T_\gamma^4, \quad  p=\tfrac{1}{3}a_{\text{B}}T_\gamma^4,
\end{equation}
where $a_{\text{B}}$ is the black body constant and $T_\gamma$ the
radiation temperature.  The equations of state~(\ref{eq:state-rad})
imply the equation of state for the pressure
$p=\tfrac{1}{3}\varepsilon$, so that, with~(\ref{eq:part-p}),
\begin{equation}
  \label{eq:state-rad-diff}
    p_n=0, \quad p_\varepsilon=\tfrac{1}{3}. 
\end{equation}
Therefore, one has from~(\ref{eq:beta-matter}),
\begin{equation}
  \label{eq:state-rad-beta}
  \beta^2=w=\tfrac{1}{3}.
\end{equation}
Using~(\ref{eq:state-rad-diff}) and~(\ref{eq:state-rad-beta}), the
perturbation equations~(\ref{subeq:final}) reduce to
\begin{subequations}
\label{subeq:final-rad}
\begin{align}
  & \ddot{\delta}_\varepsilon-H\dot{\delta}_\varepsilon-
  \left[\dfrac{1}{3}\dfrac{\nabla^2}{a^2}-
    \tfrac{2}{3}\kappa\varepsilon_\nul\right]
  \delta_\varepsilon  = 0,   \label{eq:delta-rad} \\
  & \delta_n-\tfrac{3}{4}\delta_\varepsilon = 0,  \label{eq:entropy-rad}
\end{align}
\end{subequations}
where~(80$^\ast$) has been used.  Since $p_n=0$ the right-hand side
of~(\ref{eq:delta-rad}) vanishes, implying that density perturbations
evolved adiabatically: they did not exchange heat with their
environment.  Moreover, baryons were tightly coupled to radiation
through Thomson scattering, i.e., baryons obey
$\delta_{n,\,\text{baryon}}=\tfrac{3}{4}\delta_\varepsilon$. Thus, for
baryons~(\ref{eq:entropy-rad}) is identically satisfied.  In contrast
to baryons, \textsc{cdm} is \emph{not} coupled to radiation through
Thomson scattering.  However, equation~(\ref{eq:entropy-rad}) follows
from the General Theory of Relativity, Section~8$^\ast$.  As a
consequence, equation~(\ref{eq:entropy-rad}) should be obeyed by
\emph{all} kinds of particles that interact through gravitation.  In
other words, equation~(\ref{eq:entropy-rad}) holds true for baryons as
well as \textsc{cdm}.  Since \textsc{cdm} interacts only via gravity
with baryons and radiation, the fluctuations in \textsc{cdm} are
coupled through gravitation to fluctuations in the energy density, so
that fluctuations in \textsc{cdm} also satisfy
equation~(\ref{eq:entropy-rad}).

In order to solve equation~(\ref{eq:delta-rad}) it will first be
rewritten in a form using dimensionless quantities.  The solutions of
the background equations~(\ref{subeq:einstein-flrw}) are given by
\begin{equation}
  \label{eq:exact-sol-rad}
   H\propto t^{-1}, \quad  \varepsilon_\nul\propto t^{-2}, \quad
   n_\nul\propto t^{-3/2}, \quad a\propto t^{1/2},
\end{equation}
implying that $T_{\nul\gamma}\propto a^{-1}$.  The dimensionless time
$\tau$ is defined by $\tau\coloneqq t/t_0$. Since $H\coloneqq\dot{a}/a$, one
finds that
\begin{equation}
   \dfrac{{\text{d}}^k}{c^k{\text{d}}
t^k}=\left[\dfrac{1}{ct_0}\right]^k\dfrac{{\text{d}}^k}{{\text{d}}\tau^k}=
   \left[2H(t_0)\right]^k
   \dfrac{{\text{d}}^k}{{\text{d}}\tau^k}, \quad k=1,2.  \label{dtau-n}
\end{equation}
Substituting $\delta_\varepsilon(t,\boldsymbol{x})=
\delta_\varepsilon(t,\boldsymbol{q})\exp(\text{i}\boldsymbol{q}\cdot\boldsymbol{x})$ into
equation~(\ref{eq:delta-rad}) and using~(\ref{dtau-n}) yields
\begin{equation}
  \label{eq:new-rad}
  \delta_\varepsilon^{\prime\prime}-\dfrac{1}{2\tau}\delta_\varepsilon^\prime+
  \left[\dfrac{\mu_{\text{r}}^2}{4\tau}+\dfrac{1}{2\tau^2}\right]\delta_\varepsilon=0,
\quad \tau\ge 1,
\end{equation}
where a prime denotes differentiation with respect to $\tau$. The
parameter $\mu_{\text{r}}$ is given by
\begin{equation}
     \mu_\text{r} \coloneqq
\dfrac{2\pi}{\lambda_0}\dfrac{1}{H(t_0)}\dfrac{1}{\sqrt{3}}, \quad
    \lambda_0\coloneqq\lambda a(t_0),  
\label{xi}
\end{equation} 
with $\lambda_0$ the physical scale of a perturbation at time $t_0$
($\tau=1$), and $|\boldsymbol{q}|=2\pi/\lambda$.  To solve
equation~(\ref{eq:new-rad}), replace $\tau$ by
$x\coloneqq\mu_{\text{r}}\sqrt{\tau}$. After transforming back to
$\tau$, one finds
\begin{equation}
 \delta_\varepsilon(\tau,\boldsymbol{q}) =
      \Bigl[A_1(\boldsymbol{q})\sin\left(\mu_{\text{r}}\sqrt{\tau}\right) +
A_2(\boldsymbol{q})\cos\left(\mu_{\text{r}}\sqrt{\tau}\right)\Bigr]\sqrt{\tau},
\label{nu13}
\end{equation}
where the `constants' of integration
$A_1(\boldsymbol{q})$ and $A_2(\boldsymbol{q})$ are given by
\begin{equation}
\label{subeq:C1-C2}
   A_{1\atop2}(\boldsymbol{q}) =
   \delta_\varepsilon(t_0,\boldsymbol{q}) {\sin\mu_{\text{r}}\atop\cos\mu_{\text{r}}}\mp
  \dfrac{1}{\mu_{\text{r}}}{\cos\mu_{\text{r}}\atop\sin\mu_{\text{r}}}
\left[\delta_\varepsilon(t_0,\boldsymbol{q})-\dfrac{\dot{\delta}
_\varepsilon(t_0,\boldsymbol{q})}{H(t_0)}\right].
\end{equation}
For large-scale perturbations ($\lambda\rightarrow\infty$), it follows
from~(\ref{nu13}) and~(\ref{subeq:C1-C2}) that
\begin{align}
    \delta_\varepsilon(t) =  -\left[\delta_\varepsilon(t_0)-
       \dfrac{\dot{\delta}_\varepsilon(t_0)}{H(t_0)}\right]
       \dfrac{t}{t_0}
   +\left[2\delta_\varepsilon(t_0)
    - \dfrac{\dot{\delta}_\varepsilon(t_0)}{H(t_0)}\right]
     \left(\dfrac{t}{t_0}\right)^{\tfrac{1}{2}}. \label{delta-H-rad}
\end{align}
The energy density contrast has two contributions to the growth rate,
one proportional to $t$ and one proportional to $t^{1/2}$. These two
solutions have been found, with the exception of the precise factors
of proportionality, by a large number of
authors~\citep{c15,adams-canuto1975,olson1976,c11,kolb,c12}.  If one
makes the plausible assumption that density perturbations do not
evolve when they arise, i.e.,
$\dot{\delta}_\varepsilon(t_0,\boldsymbol{q})\approx0$, then only
perturbations with $\lambda\rightarrow\infty$ will die out, as follows
from~(\ref{delta-H-rad}).  In contrast, finite-scale density
perturbations grow proportional to the square root of
time~(\ref{nu13}).  For example, $\lambda\rightarrow0$ yields
\begin{equation}\label{dc-small}
   \delta_\varepsilon(t,\boldsymbol{q}) \approx
\delta_\varepsilon(t_0,\boldsymbol{q})
    \left(\dfrac{t}{t_0}\right)^{\tfrac{1}{2}}
   \cos\left[\mu_{\text{r}}-
\mu_{\text{r}}\left(\dfrac{t}{t_0}\right)^{\tfrac{1}{2}}\right],
\end{equation}
as follows from~(\ref{nu13}) and~(\ref{subeq:C1-C2}).  Thus, the
evolution equations~(\ref{subeq:final-rad}) yield oscillating density
perturbations with an \emph{increasing} amplitude.

\subsection{Plasma Era}

The plasma era has begun at time $t_{\mathrm{eq}}$, when the
energy density of ordinary matter was equal to the energy density of
radiation,~(\ref{eq:mat-en-eq}), and ends at $t_{\mathrm{dec}}$, the
time of decoupling of matter and radiation. In the plasma era the
matter-radiation mixture can be characterised by the equations of
state (see \cite{kodama1984}, Chapter~V)
\begin{equation}
  \label{eq:eos-plasma}
  \varepsilon(n,T)=nmc^2+a_{\text{B}}T^4_\gamma, \quad
        p(n,T)=\tfrac{1}{3}a_{\text{B}}T^4_\gamma,
\end{equation}
where the contributions to the pressure of ordinary matter and
\textsc{cdm} have not been taken into account, since these
contributions are negligible with respect to the radiation energy
density.  Eliminating $T_\gamma$ from~(\ref{eq:eos-plasma}), one finds
for the equation of state for the pressure
\begin{equation}
  \label{eq:press-plasma}
  p(n,\varepsilon)=\tfrac{1}{3}(\varepsilon-nmc^2),
\end{equation}
so that with~(\ref{eq:part-p}) one gets
\begin{equation}
  \label{eq:partial-plasma}
   p_n=-\tfrac{1}{3}mc^2,  \quad p_\varepsilon=\tfrac{1}{3}.
\end{equation}
Since $p_n<0$, equation~(\ref{fir-ord}) implies that fluctuations in
the particle number density, $\delta_n$, were coupled to fluctuations
in the total energy density, $\delta_\varepsilon$, through
gravitation, irrespective of the nature of the particles.

\subsection{Era after Decoupling of Matter and Radiation}
\label{sec:mat-dom-era}

Once protons and electrons combined to yield hydrogen, the radiation
pressure was negligible, and the equations of state have become those
of a non-relativistic monatomic perfect gas with three degrees of
freedom
\begin{align}
   \label{state-mat}
  \varepsilon(n,T) = nmc^2+\tfrac{3}{2}nk_{\text{B}}T, \quad
  p(n,T) = nk_{\text{B}}T,
\end{align}
where $k_{\text{B}}$ is Boltzmann's constant, $m$ the mean particle
mass, and $T$ the temperature of the matter.  For the calculations in
this subsection it is only needed that the \textsc{cdm} particle mass
is such that for the mean particle mass $m$ one has
$mc^2\gg k_{\text{B}}T$, so that
$w\coloneqq p_\nul/\varepsilon_\nul\ll1$.  Therefore, as follows from
the background equations~(\ref{FRW3}) and~(\ref{FRW2}), one may
neglect the pressure $nk_{\text{B}}T$ and the kinetic energy density
$\tfrac{3}{2}nk_{\text{B}}T$ with respect to the rest mass energy
density $nmc^2$ in the \emph{un}perturbed universe.  However,
neglecting the pressure in the \emph{perturbed} universe yields
non-evolving density perturbations with a static gravitational field,
as is shown in Section~7$^\ast$ on the non-relativistic limit.
Consequently, it is important to take the pressure perturbations into
account.  Eliminating $T$ from~(\ref{state-mat}) yields the equation
of state for the pressure
\begin{equation}
  \label{eq:pne-eq-of-state}
  p(n,\varepsilon)=\tfrac{2}{3}(\varepsilon-nmc^2),
\end{equation}
so that with~(\ref{eq:part-p}) one has
\begin{equation}
  \label{eq:pne-dpdedpdn}
   p_n=-\tfrac{2}{3}mc^2,  \quad p_\varepsilon=\tfrac{2}{3}.
\end{equation}
Substituting $p_n$, $p_\varepsilon$, and $\varepsilon$ given
by~(\ref{state-mat}) into~(\ref{eq:beta-matter}) on finds, using that
$mc^2\gg k_{\text{B}}T$,
\begin{equation}
     \beta\approx \dfrac{v_{\text{s}}}{c}=\sqrt{\dfrac{5}{3}
       \dfrac{k_{\text{B}}T_\nul}{mc^2}}, \quad
     w\approx\dfrac{k_{\text{B}}T_\nul}{mc^2},
\label{coef-nu1}
\end{equation}
with $v_{\text{s}}$ the adiabatic speed of sound and $T_\nul$ the
matter temperature.  Using that $\beta^2\approx\tfrac{5}{3}w$ and
$w\ll1$, expression~(\ref{eq:time-w}) reduces to $\dot{w}\approx-2Hw$,
so that with $H\coloneqq\dot{a}/a$ one has $w\propto a^{-2}$.  This
implies with~(\ref{coef-nu1}) that the matter temperature decays as
\begin{equation}
  \label{eq:temp-a-2}
   T_\nul\propto a^{-2}.
\end{equation}
This, in turn, implies with~(\ref{coef-nu1}) that
$\dot{\beta}/\beta=-H$.  The system~(\ref{subeq:final}) can now be
rewritten as
\begin{subequations}
  \label{final-dust}
  \begin{align}
   & \ddot{\delta}_\varepsilon + 3H\dot{\delta}_\varepsilon-
  \left[\beta^2\dfrac{\nabla^2}{a^2}+
   \tfrac{5}{6}\kappa\varepsilon_\nul\right]
   \delta_\varepsilon=-\dfrac{2}{3}\dfrac{\nabla^2}{a^2}\left(\delta_n-\delta_\varepsilon\right),
   \label{dde-dn-de}\\
   & \dfrac{1}{c}\dfrac{{\text{d}}}{{\text{d}} t}
   \left(\delta_n-\delta_\varepsilon\right)=
   -2H\left(\delta_n-\delta_\varepsilon\right), \label{eq:dn-de}
  \end{align}
\end{subequations}
where $w\ll1$ and $\beta^2\ll1$ have been neglected with respect to
constants of order unity.  From equation~(\ref{eq:dn-de}) it follows
with $H\coloneqq\dot{a}/a$ that
\begin{equation}
  \label{eq:dn-dn-a-2}
  \delta_n-\delta_\varepsilon \propto a^{-2}.
\end{equation}
Since the system~(\ref{final-dust}) is derived from the General Theory
of Relativity, it should be obeyed by all kinds of particles which
interact through gravity, in particular baryons and \textsc{cdm}.

It will now be shown that the right-hand side of
equation~(\ref{dde-dn-de}) is proportional to the mean kinetic energy
density fluctuation of the particles of a density perturbation.  To
that end, an expression for $\varepsilon^\true_\een$ will be derived
from~(\ref{state-mat}).  Multiplying $\dot{\varepsilon}_\nul$ by
$\theta_\een/\dot{\theta}_\nul$ and subtracting the result from
$\varepsilon_\een$, one finds
\begin{equation}
  \label{eq:gauge-dep-e1}
  \varepsilon^\true_\een=n^\true_\een mc^2+\tfrac{3}{2}n^\true_\een k_{\text{B}}T_\nul+
            \tfrac{3}{2}n_\nul k_{\text{B}}T^\true_\een,
\end{equation}
where also the expressions~(60$^\ast$) and~(74$^\ast$) have been
used.  Dividing the result by $\varepsilon_\nul$,~(\ref{state-mat}),
and using that $k_{\text{B}}T_\nul\ll mc^2$, one finds
\begin{equation}
  \label{eq:dn-de-dT}
  \delta_\varepsilon\approx \delta_n+
        \dfrac{3}{2}\dfrac{k_{\text{B}}T_\nul}
        {mc^2}\delta_T,
\end{equation}
to a very good approximation.  In this expression $\delta_\varepsilon$
is the relative perturbation in the \emph{total} energy density.
Since $mc^2\gg\tfrac{3}{2}k_{\text{B}}T_\nul$, it follows from the
derivation of~(\ref{eq:dn-de-dT}) that $\delta_n$ is the relative
perturbation in the \emph{rest} energy density.  Consequently, the
second term is the fluctuation in the \emph{kinetic} energy density,
i.e., $\delta_{\text{kin}}\approx\delta_\varepsilon-\delta_n$. The
relative kinetic energy density perturbation occurs in the source term
of the evolution equation~(\ref{dde-dn-de}) and is of the same order
of magnitude as the term with~$\beta^2$ in the left-hand side.  This
is the very reason that pressure perturbations should not be
neglected in the perturbed universe.

Combining~(\ref{coef-nu1}), (\ref{eq:temp-a-2})
and~(\ref{eq:dn-dn-a-2}) one finds from~(\ref{eq:dn-de-dT}) that
$\delta_T$ is constant
\begin{equation}
  \label{eq:dT-constant}
  \delta_T(t,\boldsymbol{x})\approx\delta_T(t_0,\boldsymbol{x}),
\end{equation}
to a very good approximation, so that the kinetic energy density
fluctuation is given by
\begin{equation}
  \label{eq:kin-en}
 \delta_{\text{kin}}(t,\boldsymbol{x})\approx
  \delta_\varepsilon(t,\boldsymbol{x})-\delta_n(t,\boldsymbol{x}) \approx
  \dfrac{3}{2}\dfrac{k_{\text{B}}T_\nul(t)}
  {mc^2}\delta_T(t_0,\boldsymbol{x}).
\end{equation}
In Section~\ref{sec:pop-iii-stars} it will be shown that the kinetic
energy density fluctuation has played, in addition to gravitation, a
role in the evolution of density perturbations.

Using~(\ref{eq:pne-dpdedpdn}) and~(\ref{eq:dn-de-dT}), one finds
from~(\ref{eq:p-dia-adia})
\begin{equation}
  \label{eq:rel-press-pert}
  \delta_p \approx \tfrac{5}{3}\delta_\varepsilon+\delta_T,
\end{equation}
where $\delta_p$ is the relative pressure perturbation defined by
$\delta_p\coloneqq p^\true_\een/p_\nul$, with $p_\nul$ given
by~(\ref{state-mat}).  The term $\tfrac{5}{3}\delta_\varepsilon$ is
the adiabatic part and $\delta_T$ is the diabatic part of the relative
pressure perturbation. The factor $\tfrac{5}{3}$ is the so-called
adiabatic index for a monatomic ideal gas with three degrees of
freedom.  Thus, relative kinetic energy density perturbations give
rise to diabatic pressure fluctuations.

Finally, the perturbed entropy per particle follows
from~(\ref{eq:thermo-een}) and~(\ref{eq:dn-de-dT})
\begin{equation}
  \label{eq:heat-exchange}
  s^\true_\een \approx \tfrac{3}{2}k_{\text{B}}\delta_T.
\end{equation}
In Section~9.2$^\ast$ it has been shown that the background entropy
per particle $s_\nul$ is independent of time.  In the \emph{linear}
regime after decoupling of matter and radiation, the perturbed entropy
per particle is approximately constant, i.e.,
$\dot{s}^\true_\een\approx0$, as follows from~(\ref{eq:dT-constant})
and~(\ref{eq:heat-exchange}).  Therefore, heat exchange of a
perturbation with its environment decays proportional to the
temperature, i.e., $T_\nul s^\true_\een\propto a^{-2}$, as follows
from~(\ref{eq:temp-a-2}).

In order to solve equation~(\ref{dde-dn-de}) it will first be
rewritten in a form using dimensionless quantities.  The solutions of
the background equations~(\ref{subeq:einstein-flrw}) are given by
\begin{equation}
  \label{eq:exact-sol-mat}
   H\propto t^{-1}, \quad  \varepsilon_\nul\propto t^{-2}, \quad
   n_\nul\propto t^{-2}, \quad a\propto t^{2/3},
\end{equation}
where the kinetic energy density and pressure have been neglected with
respect to the rest mass energy density.  The dimensionless time
$\tau$ is defined by $\tau\coloneqq t/t_0$.  Using that
$H\coloneqq\dot{a}/a$, one gets
\begin{equation}
   \dfrac{{\text{d}}^k}{c^k{\text{d}}
t^k}=\left[\dfrac{1}{ct_0}\right]^k\dfrac{{\text{d}}^k}{{\text{d}}\tau^k}=
   \left[\tfrac{3}{2}H(t_0)\right]^k
   \dfrac{{\text{d}}^k}{{\text{d}}\tau^k}, \quad k=1,2.    \label{dtau-n-dust}
\end{equation}
Substituting
$\delta_\varepsilon(t,\boldsymbol{x})=\delta_\varepsilon(t,\boldsymbol{q})\exp(\text{i}\boldsymbol{q}\cdot\boldsymbol{x})$,
$\delta_n(t,\boldsymbol{x})=\delta_n(t,\boldsymbol{q})\exp(\text{i}\boldsymbol{q}\cdot\boldsymbol{x})$,~(\ref{coef-nu1})
and~(\ref{eq:kin-en}) into equations~(\ref{final-dust}) and
using~(\ref{eq:temp-a-2}) and~(\ref{dtau-n-dust}) one finds that
equations~(\ref{final-dust}) can be combined into one equation
\begin{equation}\label{eq:dust-dimless}
    \delta_\varepsilon^{\prime\prime}+\dfrac{2}{\tau}\delta_\varepsilon^\prime+
\left[\dfrac{4}{9}\dfrac{\mu_{\text{m}}^2}{\tau^{8/3}}-\dfrac{10}{9\tau^2}
\right]\delta_\varepsilon=
-\dfrac{4}{15}\dfrac{\mu^2_{\text{m}}}{\tau^{8/3}}
\delta_T(t_0,\boldsymbol{q}), \quad \tau\ge1,
\end{equation}
where a prime denotes differentiation with respect to~$\tau$. The
parameter $\mu_{\text{m}}$ is given~by
\begin{equation}\label{eq:const-mu}
\mu_{\text{m}}\coloneqq\dfrac{2\pi}{\lambda_0}\dfrac{1}{H(t_0)}\dfrac{
v_{\text{s}}(t_0)}{c},  \quad \lambda_0\coloneqq\lambda a(t_0),
\end{equation}
with $\lambda_0$ the physical scale of a perturbation at time $t_0$
i.e., $\tau=1$, and $|\boldsymbol{q}|=2\pi/\lambda$.  In
equation~(\ref{eq:dust-dimless}) four terms can be recognised that
increase or decrease the growth, i.e.,
\begin{subequations}
  \label{eq:evolution-terms}
\begin{alignat}{2}
 & \text{Expansion:}\quad +\dfrac{2}{\tau}\delta_\varepsilon^\prime, \qquad \label{eq:exp-press}
 && \text{Pressure:} \quad +\dfrac{4}{9}\dfrac{\mu_{\text{m}}^2}{\tau^{8/3}}\delta_\varepsilon, \\
 & \text{Gravity:} \quad \;\;\;\;\,-\dfrac{10}{9\tau^2}\delta_\varepsilon, \qquad \label{eq:grav-entro}
 && \text{Entropy:} \quad \,-\dfrac{4}{15}\dfrac{\mu^2_{\text{m}}}{\tau^{8/3}}\delta_T(t_0,\boldsymbol{q}).
\end{alignat}
\end{subequations}
When $\delta_\varepsilon>0$ the pressure counteracts the growth of
density perturbations and gravity increases its growth. The
expansion counteracts the growth when $\delta^\prime_\varepsilon>0$ and
increases the growth when $\delta^\prime_\varepsilon<0$.  The entropy
increases the growth if $\delta_T(t_0,\boldsymbol{q})<0$.

To derive an analytical solution of equation~(\ref{eq:dust-dimless})
replace $\tau$ by $x\coloneqq2\mu_{\text{m}}\tau^{-1/3}$. After
transforming back to $\tau$, one finds for the general solution of the
evolution equation~(\ref{eq:dust-dimless})
\begin{equation}
  \label{eq:matter-physical}
 \delta_\varepsilon(\tau,\boldsymbol{q}) =
    \Bigl[B_1(\boldsymbol{q})
      J_{+\frac{7}{2}}\bigl(2\mu_{\text{m}}\tau^{-1/3}\bigr)
          + B_2(\boldsymbol{q})J_{-\frac{7}{2}}
       \bigl(2\mu_{\text{m}}\tau^{-1/3}\bigr)\Bigr]\tau^{-1/2}
      -\dfrac{3}{5}\left[1+
        \dfrac{5}{2}\dfrac{\tau^{2/3}}{\mu_{\text{m}}^2}\right]\delta_T(t_0,\boldsymbol{q}),
\end{equation}
where $J_{\pm7/2}(x)$ are Bessel functions of the first kind and
$B_1(\boldsymbol{q})$ and $B_2(\boldsymbol{q})$ are the `constants' of integration,
calculated with the help of \cite{maxima}:
\begin{align}
\label{subeq:B1-B2}
  B_{1\atop2}(\boldsymbol{q}) =&\; \dfrac{3\sqrt{\pi}}{20\mu_{\text{m}}^{3/2}}
   \left[\bigl(4\mu_{\text{m}}^2-5\bigr){\cos2\mu_{\text{m}}\atop\sin2\mu_{\text{m}}}
    \mp10\mu_{\text{m}}{\sin2\mu_{\text{m}}\atop\cos2\mu_{\text{m}}}\right]\delta_T(t_0,\boldsymbol{q})\;+
  \nonumber \\
  & \; \dfrac{\sqrt{\pi}}{8\mu_{\text{m}}^{7/2}}\left[\bigl(8\mu_{\text{m}}^4-
     30\mu_{\text{m}}^2+15\bigr){\cos2\mu_{\text{m}}\atop\sin2\mu_{\text{m}}}
       \mp\bigl(20\mu_{\text{m}}^3-30\mu_{\text{m}}\bigr){\sin2\mu_{\text{m}}\atop\cos2\mu_{\text{m}}}\right]
  \delta_\varepsilon(t_0,\boldsymbol{q})\;+ \nonumber \\
   & \; \dfrac{\sqrt{\pi}}{8\mu_{\text{m}}^{7/2}}\left[\bigl(24\mu_{\text{m}}^2-15\bigr)
       {\cos2\mu_{\text{m}}\atop\sin2\mu_{\text{m}}}\pm
      \bigl(8\mu_{\text{m}}^3-30\mu_{\text{m}}\bigr){\sin2\mu_{\text{m}}\atop\cos2\mu_{\text{m}}}
     \right]\dfrac{\dot{\delta}_\varepsilon(t_0,\boldsymbol{q})}{H(t_0)}.
\end{align}
The particle number density contrast $\delta_n(t,\boldsymbol{q})$
follows from equation~(\ref{eq:dn-de-dT}), (\ref{eq:dT-constant})
and~(\ref{eq:matter-physical}). In~(\ref{eq:matter-physical}) the
first term, i.e., the solution of the homogeneous equation, is the
adiabatic part of a density perturbation, whereas the second term,
i.e., the particular solution, is the diabatic part.

In the large-scale limit $\lambda\rightarrow\infty$ terms with
$\nabla^2$ vanish. Therefore, the general solution of
equation~(\ref{eq:dust-dimless}) becomes
\begin{equation}
    \delta_\varepsilon(t)  =
    \dfrac{1}{7}\left[5\delta_\varepsilon(t_0)+\dfrac{2\dot{\delta}_\varepsilon(t_0)}{H(t_0)}\right]
 \left(\dfrac{t}{t_0}\right)^{\tfrac{2}{3}}
     + \dfrac{2}{7}\left[\delta_\varepsilon(t_0)-\dfrac{\dot{\delta}_\varepsilon(t_0)}{H(t_0)}\right]
      \left(\dfrac{t}{t_0}\right)^{-\tfrac{5}{3}},
                               \label{eq:new-dust-53-adiabatic-e}
\end{equation}
Thus, for large-scale perturbations the diabatic pressure
fluctuation $\delta_T(t_0,\boldsymbol{q})$ did not play a role during the
evolution: large-scale perturbations were adiabatic and evolved only
under the influence of gravity.  These perturbations were so large
that heat exchange did not play a role during their evolution in the
linear phase.  For perturbations much larger than the
\emph{relativistic Jeans scale}, i.e., the peak value at
$6.5\,\text{pc}$ in Figure~\ref{fig:collapse}, gravity alone was
insufficient to explain structure formation within
$13.79\,\text{Gyr}$, since they grow as
$\delta_\varepsilon\propto t^{2/3}$.

The solution proportional to $t^{2/3}$ is a standard result
\citep{c15,adams-canuto1975,olson1976,c11,kolb,c12}.  Since
$\delta_\varepsilon$ is gauge-invariant, the standard non-physical
gauge mode proportional to $t^{-1}$ is absent from the solution set of
the evolution equations~(\ref{final-dust}).  Instead, a physical mode
proportional to $t^{-5/3}$ is found.  Consequently,
only the growing mode of~(\ref{eq:new-dust-53-adiabatic-e}) is in
agreement with results given in the literature.

In the small-scale limit $\lambda\rightarrow0$, one finds
from~(\ref{eq:matter-physical}) and~(\ref{subeq:B1-B2})
\begin{equation}
  \delta_\varepsilon(t,\boldsymbol{q}) \approx -\tfrac{3}{5}\delta_T(t_0,\boldsymbol{q})+
     \left(\dfrac{t}{t_0}\right)^{-\tfrac{1}{3}}
     \Bigl[\delta_\varepsilon(t_0,\boldsymbol{q})+\tfrac{3}{5}\delta_T(t_0,\boldsymbol{q})\Bigr]
      \cos\left[2\mu_{\text{m}}-2\mu_{\text{m}}\left(\dfrac{t}{t_0}
        \right)^{-\tfrac{1}{3}}\right]. \label{eq:small-scale-dust}
\end{equation}
Thus, density perturbations with scales smaller than the relativistic
Jeans scale oscillated with a decaying amplitude which was smaller
than unity: these perturbations were so small that gravity was
insufficient to let perturbations grow. Heat exchange alone was not
enough for the growth of density perturbations.  Consequently,
perturbations with scales smaller than the relativistic Jeans scale
did never reach the non-linear regime.

In the next section it is shown that for density perturbations with
scales of the order of the relativistic Jeans scale, the action of
both gravity and heat exchange together may result in massive
structures several hundred million years after decoupling of matter
and radiation.

\section{Structure Formation after Decoupling
  of Matter and Radiation}
\label{sec:pop-iii-stars}

In this section it is demonstrated that the relativistic evolution
equations, which include a realistic equation of state for the
pressure $p=p(n,\varepsilon)$ yields that in the era after decoupling
of matter and radiation density perturbations may have grown fast.

Up till now it is only assumed that $mc^2\gg k_{\text{B}}T$ for
baryons and \textsc{cdm}, without specifying the mass of the baryon
and \textsc{cdm} particles.  From now on it is convenient to assume
that the mass of a \textsc{cdm} particle is of the order of magnitude
of the proton mass.

\subsection{Observable Quantities}
\label{sec:p-obs-q}

The parameter $\mu_{\text{m}}$~(\ref{eq:const-mu}) will be expressed
in observable quantities, namely the present values of the background
radiation temperature, $T_{\nul\gamma}(t_{\text{p}})$, the Hubble
parameter, $H(t_{\text{p}})$, and the
redshift at decoupling, $z(t_{\text{dec}})$.  From now on the initial
time is taken to be the time at decoupling of matter and radiation:
$t_0=t_{\text{dec}}$, so that $\tau\coloneqq t/t_{\text{dec}}$.

The redshift $z(t)$ as a function of the scale factor $a(t)$ is given
by
\begin{equation}
  \label{eq:redshift}
  z(t)=\dfrac{a(t_{\text{p}})}{a(t)}-1,
\end{equation}
where $a(t_{\text{p}})$ is the present value of the scale factor and
$z(t_{\text{p}})=0$. For a flat \textsc{flrw} universe one may take
$a(t_{\text{p}})=1$.  Using the background
solutions~(\ref{eq:exact-sol-mat}), one finds from~(\ref{eq:redshift})
\begin{subequations}
  \label{eq:handig}
  \begin{align}
  H(t)&=H(t_{\text{p}})\bigl[z(t)+1\bigr]^{3/2}, \label{handig-H} \\
  t&=t_{\text{p}}\bigl[z(t)+1\bigr]^{-3/2}, \label{handig-tijd} \\
  T_{\nul\gamma}(t)&=T_{\nul\gamma}(t_{\text{p}})\bigl[z(t)+1\bigr],
     \label{handig-temp}
\end{align}
\end{subequations}
where it is used that $T_{\nul\gamma}\propto a^{-1}$ after decoupling,
as follows from~(\ref{eq:state-rad}) and~(\ref{eq:exact-sol-rad}).

The dimensionless time $\tau\coloneqq t/t_{\text{dec}}$ can be
expressed in the redshift
\begin{equation}
  \tau =\left[\dfrac{z(t_{\text{dec}})+1}{z(t)+1}\right]^{3/2},
                     \label{tau-zt}
\end{equation}
by using that $\tau=(t/t_{\text{p}})(t_{\text{p}}/t_{\text{dec}})$ and~(\ref{handig-tijd}).

Substituting~(\ref{coef-nu1}) into~(\ref{eq:const-mu}), one gets
\begin{equation}
  \label{eq:H-dec-wmap-T-dec}
  \mu_{\text{m}}=\dfrac{2\pi}{\lambda_{\text{dec}}}\dfrac{1}{H(t_{\text{dec}})}
     \sqrt{\dfrac{5}{3}
    \dfrac{k_{\text{B}}T_{\nul}(t_{\text{dec}})}{mc^2}},
        \quad \lambda_{\text{dec}}\coloneqq\lambda a(t_{\text{dec}}),
\end{equation}
where $t_{\text{dec}}$ is the time when matter decouples from
radiation and $\lambda_{\text{dec}}$ is the physical scale of a
perturbation at time $t_{\text{dec}}$.  From~(\ref{eq:handig}) one
finds
\begin{equation}
  \label{eq:H-dec-wmap}
  \mu_{\text{m}}=\dfrac{2\pi}{\lambda_{\text{dec}}}
     \dfrac{1}{cH(t_{\text{p}})}\dfrac{1}{\bigl[z(t_{\text{dec}})+1\bigr]}
\sqrt{\dfrac{5}{3}\dfrac{k_{\text{B}}T_{\nul\gamma}(t_{\text{p}})}{m}},
\end{equation}
where it is used that for the matter temperature $T_\nul$ at
decoupling one has
$T_\nul(t_{\text{dec}})=T_{\nul\gamma}(t_{\text{dec}})$.
With~(\ref{eq:H-dec-wmap}) the parameter $\mu_{\text{m}}$ is expressed
in observable quantities.

\begin{table}[!t]
  \renewcommand{\arraystretch}{1.0}
   \caption{Planck satellite results.}
  \label{tab:planck-sat-res}
\[  \begin{array}{rcl} \hline\hline    
       z(t_{\text{dec}})&= & 1090 \\
       z(t_{\text{eq}})&=  & 3387 \\
       cH(t_{\text{p}})&= & 67.66\,\text{km}\,\text{s}^{-1}\text{Mpc}^{-1} \\
       T_{\nul\gamma}(t_{\text{p}})&= & 2.725\,\text{K} \\
       t_{\text{p}}&=  & 13.79\,\text{Gyr} \\
       \delta_{T_\gamma}(t_{\text{dec}})&= & \lesssim 10^{-5}\\
       \hline\hline 
    \end{array}
\]
\end{table}

\subsection{Initial Values from the Planck Satellite}
\label{sec:wmap}
The physical quantities measured by~\cite{2020A&A...641A...6P-verkort} and physical constants
needed in the parameter $\mu_{\text{m}}$~(\ref{eq:H-dec-wmap}) of
the evolution equation~(\ref{eq:dust-dimless}) are given in Table~\ref{tab:planck-sat-res}
and~\ref{tab:phys-const}.
Substituting these values into~(\ref{eq:H-dec-wmap}), one finds
\begin{equation}\label{eq:nu-m-lambda}
    \mu_{\text{m}}=\dfrac{16.48}{\lambda_{\text{dec}}}, \quad
\lambda_{\text{dec}} \text{ in pc},
\end{equation}
The Planck observations of the fluctuations
$\delta_{T_\gamma}(t_{\text{dec}})$ in the
background radiation temperature yield for the initial value of the
fluctuations in the energy density
\begin{equation}
  \label{eq:fluct-en}
|\delta_\varepsilon(t_{\text{dec}},\boldsymbol{q})|\lesssim10^{-5}.
\end{equation}
In addition, it is assumed that
\begin{equation}
  \label{eq:d-fluct-en-dt}
  \dot{\delta}_\varepsilon(t_{\text{dec}},\boldsymbol{q}) \approx 0,
\end{equation}
implying that during the transition from the radiation-dominated era
to the era after decoupling, perturbations in the energy density were
approximately constant with respect to time.  The initial
values~(\ref{eq:fluct-en}) and~(\ref{eq:d-fluct-en-dt}) imply the
initial values for equation~(\ref{eq:dust-dimless})
\begin{equation}
  \label{eq:init-delta-tau}
  |\delta_\varepsilon(\tau=1,\boldsymbol{q})|\lesssim 10^{-5},  \quad
  \delta^\prime_\varepsilon(\tau=1,\boldsymbol{q})=0.
\end{equation}
During the linear phase of the evolution, $\delta_n(t,\boldsymbol{q})$ follows
from~(\ref{eq:dn-de-dT}) so that the initial values
$\delta_n(t_{\text{dec}},\boldsymbol{q})$ and
$\dot{\delta}_n(t_{\text{dec}},\boldsymbol{q})$ need not be specified.

\begin{table}[!t]
     \caption{Physical constants.}
  \label{tab:phys-const}
  \renewcommand{\arraystretch}{1.0}
\[  \begin{array}{rcl}\hline\hline
      m &=&  1.6726\times10^{-27}\,\text{kg}\\
        
                   \text{pc} &=& 3.0857\times10^{16}\,\text{m}=3.2616\,\text{ly} \\
         c &=& 2.9979\times10^8\,\text{m}\,\text{s}^{-1} \\
       
          k_{\text{B}} &=& 1.3806\times10^{-23}\,\text{J}\,\text{K}^{-1}\\
                    G_{\text{N}}   &=& 6.6743\times10^{-11}\,\text{m}^3\,\text{kg}^{-1}\,\text{s}^{-2} \\
          M_\odot &=& 1.9889\times10^{30}\,\text{kg} \\
           a_{\text{B}} &=& 7.5657\times10^{-16}\,\text{J}\,\text{m}^{-3}\,\text{K}^{-4} \\ \hline\hline
  \end{array}
\]
\end{table}

\subsection{Initial Values of Diabatic Pressure Fluctuations}
\label{sec:rel-kin-pert}

The initial value of a diabatic pressure
fluctuation~$\delta_T(t_{\text{dec}},\boldsymbol{q})$,~(\ref{eq:rel-press-pert}),
just after decoupling of matter and radiation cannot be obtained
directly from observation, since its value depends on how a small area
undergoes the transition from a high pressure, radiation-dominated,
era to a low pressure, matter-dominated, era.  In
fact,~$\delta_T(t_{\text{dec}},\boldsymbol{q})$ is a quantity that
takes random values.  This section describes the circumstances of this
transition, so that a estimates of the
$\delta_T(t_{\text{dec}},\boldsymbol{q})$ can be provided. To that
end, the pressure just before decoupling is compared with the pressure
just after decoupling.

The particle number density $n_\nul(t_{\text{dec}})$ at the time of
decoupling can be calculated from its value $n_\nul(t_{\text{eq}})$ at
the end of the radiation-dominated era.  By definition, at the end of
the radiation-domination era the matter energy density $n_\nul mc^2$
was equal to the energy density of the radiation:
\begin{equation}\label{eq:mat-en-eq}
n_\nul(t_{\text{eq}})mc^2=a_{\text{B}}T_{\nul\gamma}^4(t_{\text{eq}}).
\end{equation}
Since $n_\nul\propto a^{-3}$ and $T_{\nul\gamma}\propto a^{-1}$, one
finds, using~(\ref{eq:redshift}), (\ref{handig-temp})
and~(\ref{eq:mat-en-eq}), for the particle number density
\begin{equation}
  \label{eq:M-dec-n-dec}
  n_\nul(t_{\text{dec}})=
 \dfrac{a_{\text{B}}T_{\nul\gamma}^4(t_{\text{p}})}{mc^2}
   \bigl[z(t_{\text{eq}})+1\bigr]\bigl[z(t_{\text{dec}})+1\bigr]^3.
\end{equation}
At the moment of decoupling of matter and radiation, photons could not
ionise matter any more and the two constituents fell out of thermal
equilibrium. As a consequence, the high radiation pressure
$p=\tfrac{1}{3}a_{\text{B}}T^4_\gamma$ just before decoupling did go
over into the low gas pressure $p=nk_{\text{B}}T$ after decoupling. In
fact, from~(\ref{handig-temp}) and~(\ref{eq:M-dec-n-dec}) it follows
that at decoupling one has, using also the values in
Tables~\ref{tab:planck-sat-res} and~\ref{tab:phys-const},
\begin{equation}
  \label{eq:highp-lowp}
  \dfrac{n_\nul(t_{\text{dec}})k_{\text{B}}T_\nul(t_{\text{dec}})}
    {\tfrac{1}{3}a_{\text{B}}T^4_{\nul\gamma}(t_{\text{dec}})}=
   \dfrac{3k_{\text{B}}T_{\nul\gamma}(t_{\text{p}})}{mc^2}\bigl[z(t_{\text{eq}})+1\bigr]
     \approx8.2\times10^{-10},
\end{equation}
where it is used that at the moment of decoupling the matter
temperature $T_\nul(t_{\text{dec}})$ was equal to the radiation
temperature $T_{\nul\gamma}(t_{\text{dec}})$.  The fast and chaotic
transition from a high pressure epoch to a very low pressure era may
have resulted in positive or negative relative \emph{diabatic}
pressure perturbations $\delta_T$,~(\ref{eq:rel-press-pert}), with
widely differing values due to very small fluctuations
$\delta_{\text{kin}}$,~(\ref{eq:kin-en}), in the kinetic energy
density.  Large voids were created if $\delta_T>0$.  In
Section~\ref{sec:hier-struc} it will be shown that density
perturbations which were cooler than their environments may have
collapsed fast, depending on their scales.  In fact, perturbations for
which
\begin{equation}
  \label{eq:large-pp}
      \delta_T(t_{\text{dec}},\boldsymbol{q}) \lesssim -0.005,
\end{equation}
may have resulted in primordial stars, the so-called (hypothetical)
Population~\textsc{iii} stars, and larger structures, several hundred
million years after the Big Bang.

\subsection{Structure Formation}
\label{sec:hier-struc}

\begin{figure}[!t]
\begin{center}
\includegraphics[scale=0.6]{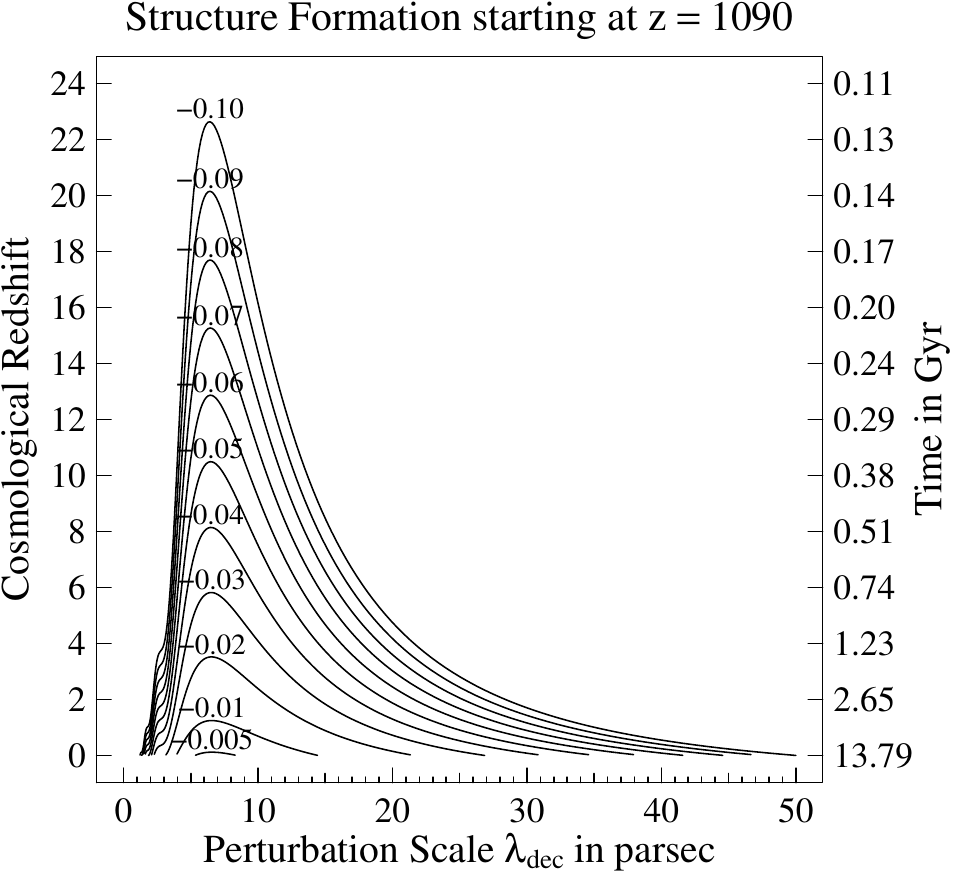}
\caption{The curves give the redshift and time, as a function of
  $\lambda_{\text{dec}}$, when a linear perturbation in the energy
  density with initial values
  $\delta_\varepsilon(t_{\text{dec}},\boldsymbol{q})\lesssim10^{-5}$ and
  $\dot{\delta}_\varepsilon(t_{\text{dec}},\boldsymbol{q})\approx0$ starting
  to grow at an initial redshift of $z(t_{\text{dec}})=1090$ has
  become non-linear, i.e., $\delta_\varepsilon(t,\boldsymbol{q})=1$.  The
  curves are labeled with the initial values of the relative
  perturbations $\delta_T(t_{\text{dec}},\boldsymbol{q})$ in the diabatic part
  of the pressure. For each curve, the Jeans scale, i.e., the peak
  value, is at $6.5\,\text{pc}$.}
\label{fig:collapse}
\end{center}
\end{figure}

In this section the evolution equation~(\ref{eq:dust-dimless}) is
solved numerically \citep{soetaert-2010,R} and the results are
summarised in Figure~\ref{fig:collapse}, which is constructed as
follows.  For each choice of $\delta_T(t_{\text{dec}},\boldsymbol{q})$
in the range $-0.005$, $-0.01$, $-0.02$, \ldots, $-0.1$
equation~(\ref{eq:dust-dimless}) is integrated for a large number of
values for the initial perturbation scale $\lambda_{\text{dec}}$ using
the initial values~(\ref{eq:init-delta-tau}). The integration starts
at $\tau=1$, i.e., at $z(t_{\text{dec}})=1090$ and will be halted if
either $z=0$, i.e., $\tau=[z(t_{\text{dec}})+1]^{3/2}$,
see~(\ref{tau-zt}), or $\delta_\varepsilon(t,\boldsymbol{q})=1$ for
$z>0$ has been reached.  One integration run yields one point on the
curve for a particular choice of the scale $\lambda_{\text{dec}}$ if
$\delta_\varepsilon(t,\boldsymbol{q})=1$ has been reached for $z>0$.
If the integration halts at $z=0$ and still
$\delta_\varepsilon(t_{\text{p}},\boldsymbol{q})<1$, then the
perturbation pertaining to that particular scale
$\lambda_{\text{dec}}$ has not yet reached its non-linear phase today,
i.e., at $t_{\text{p}}=13.79\,\text{Gyr}$. On the other hand, if the
integration is stopped at $\delta_\varepsilon(t,\boldsymbol{q})=1$ and
$z>0$, then the perturbation has become non-linear within
$13.79\,\text{Gyr}$.  Each curve denotes the time and scale for which
$\delta_\varepsilon(t,\boldsymbol{q})=1$ for a particular value of
$\delta_T(t_{\text{dec}},\boldsymbol{q})$.

The growth of a perturbation was governed by gravity as well as heat
exchange.  From Figure~\ref{fig:collapse} one may infer that the
optimal scale for growth was around $6.5\,\text{pc}$.  At this scale,
which is independent of the initial value of the diabatic pressure
perturbation $\delta_T(t_{\text{dec}},\boldsymbol{q})$,
see~(\ref{eq:p-dia-adia}) and~(\ref{eq:rel-press-pert}), heat exchange
and gravity worked together perfectly, resulting in a fast growth.
Perturbations with scales smaller than $6.5\,\text{pc}$ reached their
non-linear phase at a much later time, because their internal gravity
was weaker than for large-scale perturbations and heat exchange was
insufficient to enhance the growth. On the other hand, perturbations
with scales larger than $6.5\,\text{pc}$ exchanged heat at a slower
rate due to their large scales, so that they reached the non-linear
phase much later.  Perturbations larger than $50\,\text{pc}$ grew
proportional to $t^{2/3}$,~(\ref{eq:new-dust-53-adiabatic-e}), a
well-known result. These large-scale perturbations did not reach the
non-linear regime within $13.79\,\text{Gyr}$.

\subsubsection{Relativistic Jeans Scale}
  \label{subsubseq:rel-jeans}

Since the growth rate
decreased rapidly for perturbations with scales below
$6.5\,\text{pc}$, this scale is the \emph{relativistic} Jeans scale in
an \emph{expanding} universe at decoupling:
\begin{equation}
  \lambda_{\text{J}}(t_{\text{dec}})\coloneqq\lambda_{\text{J}}
a(t_{\text{dec}})\approx6.5\,\text{pc} = 21\,\text{ly}.
\end{equation}
The value of $\lambda_{\text{J}}(t_{\text{dec}})$ was much smaller
than the horizon size at decoupling,
$d_{\text{H}}(t_{\text{dec}})=3ct_{\text{dec}}\approx3.5\times10^5\,\text{pc}
\approx1.1\times10^6\,\text{ly}$.  The Jeans mass at decoupling,
$M_{\text{J}}(t_{\text{dec}})$, can be estimated by assuming that a
density perturbation has a spherical symmetry with diameter the
relativistic Jeans scale $\lambda_{\text{J}}(t_{\text{dec}})$. The
relativistic Jeans mass at decoupling is then given by
\begin{equation}
  \label{eq:M-dec-rel}
  M_{\text{J}}(t_{\text{dec}})=
     \dfrac{4\pi}{3}\left[\tfrac{1}{2}\lambda_{\text{J}}(t_{\text{dec}})\right]^3
     n_\nul(t_{\text{dec}})m \approx4.3\times10^3 M_\odot,
\end{equation}
where~(\ref{eq:M-dec-n-dec}) and the values in
Table~\ref{tab:planck-sat-res} and~\ref{tab:phys-const} have been
used.

\subsubsection{Classical Jeans Scale}
  \label{subsubseq:class-jeans}

The usual Jeans scale in the literature follows from the standard
equation for the gauge-dependent contrast function
$\delta\coloneqq\varepsilon_\een/\varepsilon_\nul$, i.e.,
\begin{equation}\label{eq:delta-standard}
  \ddot{\delta} + 2H\dot{\delta}-
  \left[\beta^2\frac{\nabla^2}{a^2}+
   \tfrac{1}{2}\kappa\varepsilon_\nul(1+w)(1+3w)\right]
   \delta =0.
\end{equation}
by taking $H=0$, $w=0$ and
$\beta=v_\text{s}/c$, see equations (1.1) and (2.11)--(2.13) in the
classic article of \cite{10.1093/mnras/117.1.104}.  At decoupling its
value is given by
\begin{equation}
  \label{eq:classical-Jeans-scale}
  \lambda_{\text{J},H=0}(t_{\text{dec}})=\left(\dfrac{\pi}{G_{\text{N}}\rho}
      \dfrac{\text{d}p}{\text{d}\rho}\right)^{\tfrac{1}{2}}=\left(\dfrac{\pi v^2_{\text{s}}(t_{\text{dec}})}
          {G_{\text{N}}n_\nul(t_{\text{dec}})m}\right)^{\tfrac{1}{2}}
      \approx31\,\text{pc}=103\,\text{ly},
\end{equation}
where~(\ref{eq:M-dec-n-dec}) and the values in
Tables~\ref{tab:planck-sat-res} and~\ref{tab:phys-const} are used. The
corresponding Jeans mass is
\begin{equation}
  \label{eq:M-dec-bonnor}
  M_{\text{J},H=0}(t_{\text{dec}})=
     \dfrac{4\pi}{3}\left[\tfrac{1}{2}\lambda_{\text{J},H=0}(t_{\text{dec}})\right]^3
     n_\nul(t_{\text{dec}})m=4.9\times10^5 M_\odot.
\end{equation}
As a result of the expansion, the relativistic Jeans
mass~(\ref{eq:M-dec-rel}) is a factor $114$ smaller than the usual
Jeans mass.

It will be shown in the appendix that
equation~(\ref{eq:delta-standard}) is questionable, whereas taking
$H=0$ is simply wrong because this violates the Friedmann
equation~(\ref{FRW3}).  Consequently, one may not apply the Jeans
theory to a non-static universe.

\subsection{Growth Rates}
\label{subsec:growth-rates}

The growth rate $\delta^\prime_\varepsilon$ in
equation~(\ref{eq:dust-dimless}) is expressed in $\tau^{-1}$, where
$\tau\coloneqq t/t_{\text{dec}}$ is the dimensionless time. To express
the growth rate in $\text{Gyr}^{-1}$, one needs~(\ref{dtau-n-dust}),
i.e.,
\begin{equation}
  \label{eq:tau-to-time}
  c\dot{\delta}_\varepsilon(t,\boldsymbol{q})=
  \dfrac{\text{d}\delta_\varepsilon(t,\boldsymbol{q})}{\text{d}t}=
  \tfrac{3}{2}cH(t_{\text{dec}})\delta^\prime_\varepsilon(\tau,\boldsymbol{q})\;\; [\text{s}^{-1}]
  =0.1038\,\delta^\prime_\varepsilon(\tau,\boldsymbol{q})\;\; [\text{Gyr}^{-1}],
\end{equation}
where~(\ref{handig-H}) and Tables~\ref{tab:planck-sat-res}
and~\ref{tab:phys-const} are used.

\begin{figure}[!t]
\begin{center}
\includegraphics[scale=0.6]{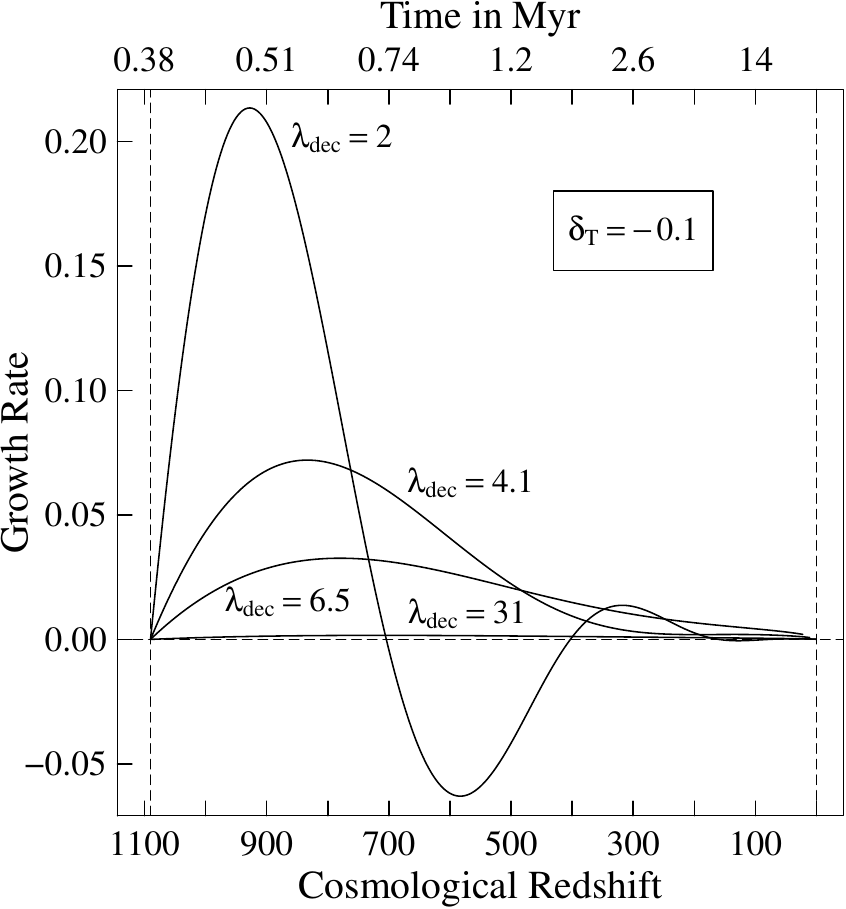}
\caption{The curves give the growth rates $\delta^\prime_\varepsilon$,
  expressed in $\tau^{-1}$, with initial value
  $\delta^\prime_\varepsilon(t_{\text{dec}},\boldsymbol{q})=0$, as
  function of the redshift~$z$, or time in million of years. The
  evolution of perturbations started at $z=1090$.  The initial
  dimensions $\lambda_{\text{dec}}$ of the perturbations are measured
  in parsec.}
\label{fig:growth-rates}
\end{center}
\end{figure}
In Figure~\ref{fig:growth-rates} the growth rates are shown for
$\delta_T(t_{\text{dec}},\boldsymbol{q})=-0.1$ for four different
values of $\lambda_{\text{dec}}$.  Since the initial
values~(\ref{eq:init-delta-tau}) were very small, the entropy in
equation~(\ref{eq:dust-dimless}) had to be positive i.e.,
$\delta_T<0$, to start the growth of a density perturbation after
decoupling.  A negative diabatic pressure fluctuation $\delta_T<0$ and
a small adiabatic pressure fluctuation
$\tfrac{5}{3}\delta_\varepsilon(t_{\text{dec}},
\boldsymbol{q})<10^{-5}$ implied that the total pressure
fluctuation~(\ref{eq:rel-press-pert}) was negative $\delta_p<0$, so
that a density perturbation could have started to grow.  The largest
initial growth rate occurred for perturbations which were smaller than
$6.5\,\text{pc}$, the peak value in Figure~\ref{fig:collapse}.  The
smallest perturbation which had become non-linear within
$13.79\,\text{Gyr}$ had a dimension of
$\lambda_{\text{dec}}=2\,\text{pc}$, as can be seen in
Figure~\ref{fig:collapse}.  All perturbations with initial dimensions
$2\,{\text{pc}}\le\lambda_{\text{dec}}<4.1\,\text{pc}$ oscillated
towards the non-linear phase within $13.79\,\text{Gyr}$.  These
perturbations grew initially so fast that after a period of time the
pressure in equation~(\ref{eq:dust-dimless}) had become larger than
the entropy, so that the growth rate decreased rather fast and had
become over time negative and the density perturbation started to
decay.  In the meantime, the density perturbation $\delta_\varepsilon$
had grown so that gravity in~(\ref{eq:dust-dimless}) did come into
play.  When the expansion term had become negative, i.e., when
$\delta^\prime_\varepsilon<0$, and since the pressure decayed faster
than the expansion term the expansion had taken over and the growth
rate increased again until the next, lower, peak in the growth rate.
For all perturbations the growth rate started to decrease when
$\delta_p>0$, i.e., when $\delta_\varepsilon>-\tfrac{3}{5}\delta_T$,
as follows from~(\ref{eq:rel-press-pert}).  In this case, the pressure
in equation~(\ref{eq:dust-dimless}) had become larger than the
entropy.  However, gravity decayed slower than the pressure, so that
gravity had become gradually stronger than the counteracting pressure
and a density perturbation could continue to grow.  After about
$15$~million years, the pressure and entropy no longer played a
significant role so that the most turbulent growth phase of a density
perturbation was over and a phase of slow and steady growth by its own
gravity towards the non-linear regime had begun.

\section{Conclusion}
\label{sec:conclusion}

The new perturbation theory developed for density perturbations in
closed, flat and open \textsc{flrw} universes \citep{miedema:2014ada}
has been applied to a flat \textsc{flrw} universe in the
radiation-dominated era and in the epoch after decoupling of matter
and radiation. A first result is that in the radiation-dominated era
oscillating density perturbations with an \emph{increasing} amplitude
proportional to $t^{1/2}$ are found, whereas the standard perturbation
equation~(\ref{eq:delta-standard}) with $w=0$ yields oscillating density
perturbations with a \emph{constant} amplitude.  This difference is
due to the fact that in the perturbation equations~(\ref{subeq:final})
the divergence $\vartheta_\een$ of the spatial part of the fluid
four-velocity is taken into account, whereas $\vartheta_\een$ is
missing in the standard equation.

In the radiation-dominated era and the plasma era baryons were tightly
coupled to radiation via Thomson scattering until decoupling.  A
second result is that \textsc{cdm} was also tightly coupled to
radiation, not through Thomson scattering, but through
\emph{gravitation}.  The fact that \textsc{cdm} and radiation are
coupled by gravitation follows from the entropy evolution
equation~(\ref{fir-ord}) since $p_n\le0$ throughout the history of the
universe as has been shown in Section~\ref{sec:analytic-sol}.  This
implies that before decoupling perturbations in \textsc{cdm} have
contracted as fast as perturbations in the baryon density. As a
consequence, \textsc{cdm} could not have triggered structure formation
after decoupling.

From observations \citep{2011ApJS..192...18K} of the Cosmic Microwave
Background it follows that perturbations were adiabatic at the moment
of decoupling, and density fluctuations $\delta_\varepsilon$ and
$\delta_n$ were of the order of $10^{-5}$ or less.  Since the growth
rate of \emph{adiabatic} perturbations in the era after decoupling was
too small to explain structure in the universe, there must have been,
in addition to gravitation, some mechanism other than \textsc{cdm}
which has enhanced the growth rate sufficiently to form the first
stars from small density perturbations.  A final result of the present
study is that it has been demonstrated that after decoupling such a
mechanism did indeed exist in the early universe.

At the moment of decoupling of matter and radiation, photons could not
ionise matter any more and the two constituents fell out of thermal
equilibrium.  As a consequence, the pressure dropped from a very high
radiation pressure just before decoupling to a very low gas pressure
after decoupling.  This fast and chaotic transition from a high
pressure epoch to a very low pressure era may have resulted in large
relative \emph{diabatic} pressure perturbations due to very small
fluctuations in the kinetic energy density.  It is found that the
growth of a density perturbation has not only been governed by
gravitation, but also by heat exchange of a perturbation with its
environment.  The growth rate depended strongly on the scale of a
perturbation.  For perturbations with a scale of
$6.5\,\text{pc}\approx 21\,\text{ly}$, the peak value in
Figure~\ref{fig:collapse}, gravity and heat exchange worked perfectly
together, resulting in a fast growth rate.  Perturbations larger than
this scale reached, despite their stronger gravitational field, their
non-linear phase at a later time since heat exchange was low due to
their larger scales.  On the other hand, for perturbations with scales
much smaller than $6.5\,\text{pc}$ gravity was too weak and heat
exchange was not sufficient to let perturbations grow.  Therefore,
density perturbations with scales much smaller than $6.5\,\text{pc}$
did not reach the non-linear regime within $13.79\,\text{Gyr}$, the
age of the universe.  Since there was a sharp decline in growth rate
below a scale of $6.5\,\text{pc}$, this scale will be called the
\emph{relativistic Jeans scale}.

The conclusion of the present article is that the
$\Lambda\textsc{cdm}$ model of the universe and its evolution
equations for density perturbations~(\ref{subeq:final}) explain the
so-called hypothetical Population~\textsc{iii} stars and larger
structures in the universe, which came into existence several hundreds
of million years after the Big Bang as has been observed by
\cite{2015arXiv150300002W,2015ApJ...808..139S}.

\appendix
\numberwithin{equation}{section}

\section{Why the Standard Equation
  is inadequate to study Density Perturbations}
\label{sec:stand-theory}

The standard evolution equation for relative density perturbations
$\delta(t,\boldsymbol{x})$ in a flat, $R_\nul=0$, \textsc{flrw}
universe with vanishing cosmological constant, $\Lambda=0$, is given
by~(\ref{eq:delta-standard}).  In the radiation-dominated universe one
has $\beta^2=\tfrac{1}{3}$ and $w=\tfrac{1}{3}$ and this equation is
identical to the relativistic equation~(15.10.57) in the textbook of
\cite{c8}. In order to arrive at equation~(\ref{eq:delta-standard})
\cite{c8} applies the large-scale limit
$\boldsymbol{q}\rightarrow\boldsymbol{0}$, implying that
$\nabla^2\delta\rightarrow0$, so that the pressure term with $\beta^2$
does not occur.  However, $\boldsymbol{q}\rightarrow\boldsymbol{0}$
yields, in addition to $\nabla^2\delta\rightarrow0$, also
$\boldsymbol{q}\cdot\boldsymbol{U}_1\rightarrow0$, which is equivalent
to $\vartheta_\een\coloneqq u^k_{\een|k}\rightarrow0$, so that there
is no fluid flow at all.  In Section~7$^\ast$ it is shown that
$\vartheta_\een\rightarrow0$ yields the non-relativistic
limit. Therefore, the limit $\vartheta_\een\rightarrow0$ is not
allowed.  In the epoch after decoupling of matter and radiation
$\beta$ is given by~(\ref{coef-nu1}), so that
$w\approx\tfrac{3}{5}\beta^2\ll1$.  In this
case~(\ref{eq:delta-standard}) is identical to equation~(15.9.23) of
\cite{c8} which has been derived using the Newtonian Theory of
Gravity.

In this section it will be shown that the standard relativistic
equation~(\ref{eq:delta-standard}) is inadequate to study the
evolution of density perturbations in the universe.  To that end,
equation~(\ref{eq:delta-standard}) will be derived from the full set
of linearized Einstein equations and conservation laws~(32$^\ast$),
or, equivalently,~(62$^\ast$) in the companion article.

Since the source term of~(\ref{eq:delta-standard}) is zero, this
equation describes \emph{adiabatic} perturbations, Section~9.3$^\ast$,
which evolve only under the influence of their own gravitational
field.  Therefore, the equation of state is given by
$p=p(\varepsilon)$. This implies that $p_n=0$, so that
$\dot{p}_\nul=p_\varepsilon\dot{\varepsilon}_\nul$ and
$p_\een=p_\varepsilon\varepsilon_\een$.  Consequently, the evolution
equations for the background particle number density
$n_\nul$,~(\ref{FRW2a}), and its first-order perturbation
$n_\een$,~(62b$^\ast$), need not be considered.
From~(\ref{eq:beta-matter}) one finds that $p_\varepsilon=\beta^2$ so
that $p_\een=\beta^2\varepsilon_\een$.  Using the definition
$\delta\coloneqq\varepsilon_\een/\varepsilon_\nul$ and $R_\nul=0$
equations~(62$^\ast$) for scalar perturbations can be written in the
form
\begin{subequations}
\label{subeq:pertub-gi}
\begin{align}
  \label{eq:dot-delta}
  \dot{\delta} + 3H\delta\Bigl[\beta^2+\tfrac{1}{2}(1-w)\Bigr]+
     (1 + w)\left[\vartheta_\een + \dfrac{R_\een}{4H}\right]&=0, \\
  \dot{\vartheta}_\een + H(2-3\beta^2)\vartheta_\een +
        \frac{\beta^2}{1+w}
      \dfrac{\nabla^2\delta}{a^2} &= 0,
\label{FRW5gi}\\
  \dot{R}_\een + 2HR_\een - 
     2\kappa\varepsilon_\nul(1 + w)\vartheta_\een & =0, \label{FRW6gi}
\end{align}
\end{subequations}
where also the background equations~(\ref{subeq:einstein-flrw}) have
been used.  These equations include the complete set of Einstein
equations and conservation laws for scalar perturbations as follows
from their derivation in the companion article \citep{miedema:2014ada}
in Sections~$5^\ast$ and~$8^\ast$. In fact,
equation~(\ref{eq:dot-delta}) is the energy density conservation law
combined with the energy density constraint
equation. Equation~(\ref{FRW5gi}) is the momentum conservation law or
continuity equation, and~(\ref{FRW6gi}) is the momentum constraint
equation.  As is shown in Section~$5^\ast$, the dynamical equations
$G^i_{\een j}$ are not needed, since they are automatically fulfilled
by the system~(\ref{subeq:pertub-gi}).

Differentiating~(\ref{eq:dot-delta}) with respect to time and
eliminating the time-derivatives of $H$, $\varepsilon_\nul$,
$\vartheta_\een$ and $R_\een$ with the help of the system of
equations~(\ref{subeq:einstein-flrw}) and perturbation
equations~(\ref{FRW5gi}) and~(\ref{FRW6gi}), respectively, and,
subsequently, eliminating $R_\een$ with the help
of~(\ref{eq:dot-delta}), one finds, using \cite{maxima}, that the set
of equations~(\ref{subeq:pertub-gi}) can be recast in the form
\begin{subequations}
\label{subeq:algemene-vergelijkingen}
\begin{align}
 & \ddot{\delta}+2H\dot{\delta}\Bigl[1+3\beta^2-3w\Bigr]-\biggl[\beta^2\dfrac{\nabla^2}{a^2}+
   \tfrac{1}{2}\kappa\varepsilon_\nul\Bigl((1+w)(1+3w) \Bigr. \Bigr. \nonumber \\
 &    \Bigl. \biggl. \phantom{\ddot{\delta}+2H\dot{\delta}\Bigl[1}
        +4w-6w^2+12\beta^2w-4\beta^2-6\beta^4 \Bigr)-6\beta\dot{\beta}H\biggr]\delta  
    =-3H\beta^2(1+w)\vartheta_\een, \label{eq:algemeen-p-e} \\
 &  \dot{\vartheta}_\een+H(2-3\beta^2)\vartheta_\een+
           \dfrac{\beta^2}{1+w}\dfrac{\nabla^2\delta}{a^2}=0, \label{eq:continuity-equation}
\end{align}
\end{subequations}
where $\dot{w}$ has been eliminated using~(\ref{eq:time-w}).  The
system~(\ref{subeq:algemene-vergelijkingen}) consists of two
relativistic equations for the unknown quantities $\delta$ and
$\vartheta_\een$.  Thus, the relativistic perturbation
equations~(62$^\ast$) for open, flat or closed \textsc{flrw} universes
and a general equation of state for the pressure $p=p(n,\varepsilon)$
reduce for a flat \textsc{flrw} universe and a barotropic equation of
state $p=p(\varepsilon)$ to the
system~(\ref{subeq:algemene-vergelijkingen}). Therefore, the
system~(\ref{subeq:algemene-vergelijkingen}) describes adiabatic
density perturbations, notwithstanding the non-zero source term
in~(\ref{eq:algemeen-p-e}).

The gauge modes~(42$^\ast$)
\begin{equation}
  \label{eq:gauge-modes-standard}
  \hat{\delta}(t,\boldsymbol{x})\coloneqq
     \dfrac{\psi(\boldsymbol{x})\dot{\varepsilon}_\nul(t)}{\varepsilon_\nul(t)}=
     -3H(t)\psi(\boldsymbol{x})\bigl[1+w(t)\bigr],
  \quad \hat{\vartheta}_\een(t,\boldsymbol{x})=
        -\dfrac{\nabla^2\psi(\boldsymbol{x})}{a^2(t)},
\end{equation}
are, for all scales, solutions of
equations~(\ref{subeq:algemene-vergelijkingen}), with $\dot{w}$ given
by~(\ref{eq:time-w}), as can be verified by substitution.

The usual way to solve the
system~(\ref{subeq:algemene-vergelijkingen}), is to impose
\emph{physical} initial conditions $\delta^\true(t_0,\boldsymbol{x})$,
$\dot{\delta}^\true(t_0,\boldsymbol{x})$ and
$\vartheta_\een^\true(t_0,\boldsymbol{x})$, so that the initial values
are written as linear combinations of the constants of integration.
This standard procedure is not feasible for the cosmological
perturbation equations~(\ref{subeq:algemene-vergelijkingen}), since
the gauge modes~(\ref{eq:gauge-modes-standard}) are solutions of the
system~(\ref{subeq:algemene-vergelijkingen}), so that some integration
constants are in fact gauge modes, which cannot be fixed by physical
initial conditions.  Consequently, the presence of gauge modes in the
linearized Einstein equations and conservation laws hinders the
formulation of a correct perturbation theory.

A striking difference between equations~(\ref{eq:delta-standard})
and~(\ref{eq:algemeen-p-e}) is that the right-hand side of the
standard equation~(\ref{eq:delta-standard}) is zero, whereas the
right-hand side of equation~(\ref{eq:algemeen-p-e})
contains~$\vartheta_\een$.  The divergence~$\vartheta_\een$ is an
intrinsic property of a density perturbation and thus should be
incorporated in the left-hand side of a perturbation equation as is
the case in equation~(\ref{sec-ord}). The fact that
$\vartheta_\een\coloneqq u^k_{\een|k}$ is indeed part of the left-hand
side of equation~(\ref{sec-ord}) follows from its derivation in the
appendix of \cite{miedema:2014ada}.  The source term in
equation~(\ref{eq:algemeen-p-e}) displaces the actual source term,
namely the entropy.  In the next subsections it will be shown, using
the system~(\ref{subeq:algemene-vergelijkingen}) why the
three-divergence, $\vartheta_\een\coloneqq u^k_{\een|k}$, of the
spatial part of the fluid velocity should not be neglected and why,
after decoupling of matter and radiation, the entropy is of the utmost
importance.

\subsection{Radiation-dominated Era}
\label{subsec:rad-dom-ijk}

In this era, the pressure is given by a linear barotropic equation of
state $p=\tfrac{1}{3}\varepsilon$, so that $p_n=0$ and
$p_\varepsilon=\tfrac{1}{3}$.  From~(\ref{eq:beta-matter}), one finds
that $\beta^2=\tfrac{1}{3}$, so that~(\ref{eq:time-w}) is identically
satisfied.  In this case
equations~(\ref{subeq:algemene-vergelijkingen}) reduce to
\begin{subequations}
\label{subeq:standard}
\begin{align}
 & \ddot{\delta} + 2H\dot{\delta}-
  \left[\dfrac{1}{3}\frac{\nabla^2}{a^2}+
   \tfrac{4}{3}\kappa\varepsilon_\nul\right]
   \delta =-\tfrac{4}{3}H\vartheta_\een, 
        \label{eq:delta-standard-genrel}  \\
 &  \dot{\vartheta}_\een+H\vartheta_\een+
           \dfrac{1}{4}\dfrac{\nabla^2\delta}{a^2}=0.
  \label{eq:continuity}
\end{align}
\end{subequations}
The gauge modes~(\ref{eq:gauge-modes-standard}) are solutions of the
system~(\ref{subeq:standard}) for $w=\tfrac{1}{3}$.  For large-scale
perturbations one has $\nabla^2\delta^\true\rightarrow0$, so that the solution of the
system~(\ref{subeq:standard}) is
\begin{subequations}
  \label{eq:cont-eq-sol-phys-gauge}
  \begin{align}
    \delta & =(c_1t - 2\psi t^{-1}) + \tfrac{9}{8}t^{1/2}, \label{eq:cont-eq-sol-phys-gauge-A} \\
    \vartheta_\een & =-\dfrac{\nabla^2\psi}{a^2}+\tfrac{9}{8}t^{-1/2},
                     \label{eq:cont-eq-sol-phys-gauge-B}
  \end{align}
\end{subequations}
where it is used that $3H^2=\kappa\varepsilon_\nul$ and $H=1/2t$.  The
expression between brackets in~(\ref{eq:cont-eq-sol-phys-gauge-A}) is
a solution of the homogeneous part of
equation~(\ref{eq:delta-standard-genrel}), whereas the particular
solution~$\tfrac{9}{8}t^{1/2}$ comes from the physical part
of~$\vartheta_\een$.  Since the physical part of $\vartheta_\een$
in~(\ref{eq:cont-eq-sol-phys-gauge-B}) is \emph{positive}, the fluid
flows out of a density perturbation, so that a density perturbation
will be diluted, in accordance the remark below
expression~(\ref{delta-H-rad}).

The solution~(\ref{eq:cont-eq-sol-phys-gauge-A}) will be compared
with the solution of the system~(\ref{subeq:final-rad}) for
large-scale perturbations, which reads
\begin{equation}
  \label{eq:rad-large-scale-c1c2}
  \delta_\varepsilon(t)=d_1t+d_2t^{1/2},
\end{equation}
which does not contain the gauge mode.  The particular solution of
equation~(\ref{eq:delta-standard-genrel}) is a solution of the
homogeneous equation~(\ref{eq:delta-rad}) since~$\vartheta_\een$ is
part of equation~(\ref{eq:delta-rad}).  Hence the differences in
appearance between~(\ref{eq:delta-rad})
and~(\ref{eq:delta-standard-genrel}).
Comparing~(\ref{eq:cont-eq-sol-phys-gauge-A})
and~(\ref{eq:rad-large-scale-c1c2}) it must be concluded that the
gauge mode has displaced the physical solution from its proper place,
since~$\vartheta_\een$ occurs in the source term
of~(\ref{eq:delta-standard-genrel}).

The fact that $\vartheta_\een$ is absent in~(\ref{eq:delta-standard})
is detrimental to our view of cosmological density perturbations in
the radiation-dominated universe.  Since $\nabla^2\delta^\true$ could
have been large for small-scale perturbations, it may have a large
influence on $\vartheta_\een$ and this may have, in turn, a major
impact on the evolution of $\delta^\true$.  That is
why~(\ref{eq:delta-rad}) yields oscillating density
perturbations~(\ref{nu13}) with an \emph{increasing} amplitude,
instead of a \emph{constant} amplitude as follows
from~(\ref{eq:delta-standard}).

The solutions of the standard equation~(\ref{eq:delta-standard}) and
the system~(\ref{subeq:standard}) contain gauge modes.  Moreover,
equation~(\ref{eq:delta-standard}) is incomplete.  Therefore,
both~(\ref{eq:delta-standard}) and~(\ref{subeq:standard}) are
inadequate to study small-scale density perturbations in the
radiation-dominated universe.

\subsection{Era after Decoupling of Matter and
  Radiation}

In this era, the equation of state for the pressure is, according to
thermodynamics, given by~(\ref{eq:pne-eq-of-state}), so that in this
case one should have $p=p(n,\varepsilon)$. In the literature one takes
$p=0$ in the background as well as in the perturbed universe.
Obviously, the pressure can be neglected in the background universe
since $w\ll1$ in equation~(\ref{FRW2}).  However, the pressure in the
perturbed universe should not be neglected, since $p=0$ yields the
non-relativistic limit, as is shown in Section~7$^\ast$.  Therefore,
the equations of state~(\ref{state-mat}) will be used to
derive~(\ref{eq:delta-standard}).  From~(\ref{coef-nu1}) it follows
that $w\approx\tfrac{3}{5}\beta^2\ll1$, so that
with~(\ref{eq:temp-a-2}) one finds $\dot{\beta}/\beta=-H$.  Using that
$3H^2=\kappa\varepsilon_\nul$ one gets
$6\beta\dot{\beta}H=-2\kappa\varepsilon_\nul\beta^2$.  Substituting
the latter expression into~(\ref{eq:algemeen-p-e}) and neglecting $w$
and $\beta^2$ with respect to constants of order unity, the
system~(\ref{subeq:algemene-vergelijkingen}) reduces to
\begin{subequations}
\label{subeq:algemene-vergelijkingen-mat-dom}
\begin{align}
 & \ddot{\delta}+2H\dot{\delta}-\left[\beta^2\dfrac{\nabla^2}{a^2}+
   \tfrac{1}{2}\kappa\varepsilon_\nul\right]\delta=-3H\beta^2\vartheta_\een,
      \label{eq:alg-mat-dom-delta} \\
 &  \dot{\vartheta}_\een+2H\vartheta_\een+
           \beta^2\dfrac{\nabla^2\delta}{a^2}=0. \label{eq:cont-decoupling}
\end{align}
\end{subequations}
The gauge modes~(\ref{eq:gauge-modes-standard}) are solutions of the
system~(\ref{subeq:algemene-vergelijkingen-mat-dom}) for $w\ll1$ and
$\nabla^2\psi=0$, as can be verified by substitution.  Consequently,
for the system~(\ref{subeq:algemene-vergelijkingen-mat-dom}) $\psi$ is
an arbitrary infinitesimal constant $C$. This implies that
$\vartheta_\een=\vartheta^\true_\een$ is a physical quantity, since
its gauge mode
$\hat{\vartheta}_\een$,~(\ref{eq:gauge-modes-standard}), vanishes
identically.  Consequently, for large-scale perturbations the solution
of the system~(\ref{subeq:algemene-vergelijkingen-mat-dom}) is
\begin{subequations}
  \label{subeq:sol-74ab}
  \begin{align}
    \delta & =(c_1t^{2/3}-2Ct^{-1})+\tfrac{7}{9}t^{-5/3}, \label{eq:sol-74c1c2} \\
     \vartheta_\een^\true & =-\tfrac{7}{9}t^{-4/3},
  \end{align}
\end{subequations}
where it is used that $3H^2=\kappa\varepsilon_\nul$, $H=2/3t$, and
$\beta^2\propto t^{-4/3}$ as follows from~(\ref{coef-nu1})
and~(\ref{eq:temp-a-2}). The expression between brackets
in~(\ref{eq:sol-74c1c2}) is a solution of the homogeneous part of
equation~(\ref{eq:alg-mat-dom-delta}), whereas the particular
solution~$\tfrac{7}{9}t^{-5/3}$ comes from~$\vartheta^\true_\een$.
The fact that $\vartheta^\true_\een<0$, implies that the fluid flow is
towards the density perturbation, so that the density perturbation
grows~(\ref{eq:new-dust-53-adiabatic-e}).

Since $\psi=C$, the gauge transformation~(2$^\ast$) and~(4$^\ast$)
reduces to the residual gauge transformation~(51$^\ast$) in the
non-relativistic limit, i.e.,
\begin{equation}
\label{eq:gauge-trans-newt}
x^0 \rightarrow x^{\prime0}= x^0 - C, \quad
        x^i \rightarrow x^{\prime i}= x^i-\chi^i(\boldsymbol{x}).
\end{equation}
This is to be expected, since a cosmological fluid for which $w\ll1$
and $\beta^2\ll1$ can be described by non-relativistic equations of
state~(\ref{state-mat}).

The solution~(\ref{eq:sol-74c1c2}) will be compared with the solution
of the system~(\ref{final-dust}) for large-scale perturbations, which
reads
\begin{equation}
  \label{eq:after-mat-rad-c1c2}
  \delta_\varepsilon(t)=d_1t^{2/3}+d_2t^{-5/3}.
\end{equation}
This solution does not contain the gauge mode.  The particular
solution of~(\ref{eq:alg-mat-dom-delta}) is a solution of the
\emph{homogeneous} part of equation~(\ref{dde-dn-de})
since~$\vartheta_\een$ is part of the left-hand side of
equation~(\ref{dde-dn-de}).  Hence the differences in appearance
between the left-hand side of~(\ref{dde-dn-de})
and~(\ref{eq:alg-mat-dom-delta}).  Comparing~(\ref{eq:sol-74c1c2})
and~(\ref{eq:after-mat-rad-c1c2}) it must be concluded that the gauge
mode has displaced the physical solution from its proper place, due to
the fact that~$\vartheta_\een$ occurs in the source term
of~(\ref{eq:alg-mat-dom-delta}).

Just as in the radiation-dominated era, the standard
equation~(\ref{eq:delta-standard}) lacks the quantity
$\vartheta^\true_\een$ in its source term.  Although
$\nabla^2\delta^\true$ could have been large for small-scale density
perturbations in the early universe the absence of
$\vartheta^\true_\een$ is not as harmful as it is in the
radiation-dominated phase: due to the smallness of $\beta^2$ and the
non-relativistic particle velocities after decoupling, the impact of
$\vartheta^\true_\een$ on the evolution of $\delta^\true$ is low, but
non-zero.  This explains why both~(\ref{eq:delta-standard}) and the
\emph{homogeneous} part of~(\ref{dde-dn-de}) yield perturbations which
grow to slow to explain structure formation after decoupling as can be
seen from the analytical solution~(\ref{eq:matter-physical}) with
$\delta_T=0$.

The usual perturbation theory for density perturbations in the era
after decoupling does not take the entropy and its evolution equation
into account. This is a serious flaw, since the entropy, which act as
a source for the growth of density perturbations, plays an important
role in structure formation.  Therefore,
equations~(\ref{eq:delta-standard})
and~(\ref{subeq:algemene-vergelijkingen-mat-dom}) are incomplete and,
as a consequence, do not explain structure formation in the universe.
In contrast, the perturbation theory developed in
\cite{miedema:2014ada} yields for equations of state~(\ref{state-mat})
the system~(\ref{final-dust}).  The entropy term in
equation~(\ref{dde-dn-de}) is of the same order of magnitude as the
pressure term, as follows from equation~(\ref{eq:dust-dimless}).  The
calculations in Section~\ref{sec:pop-iii-stars} show that the entropy
term has in the first $15$~million years after the Big Bang a major
influence on the growth of density perturbations.  This growth is
sufficient to explain structure formation after decoupling of matter
and radiation.

\subsection{Dust-filled Universe}
\label{subsec:rel-stand-eq}

Using the system~(\ref{subeq:algemene-vergelijkingen}) it will now be
shown that the standard equation~(\ref{eq:delta-standard}) is not
valid in a universe filled with a pressure-less fluid, also known as
`dust'.  In a dust-filled universe the pressure vanishes, so that
$\beta=0$ and $w=0$.  The system~(\ref{subeq:algemene-vergelijkingen})
reduces~to
\begin{subequations}
   \label{eq:lege-vergelijking}
  \begin{align}
    & \ddot{\delta}+2H\dot{\delta}-
         \tfrac{1}{2}\kappa\varepsilon_\nul\delta=0, \label{eq:lege-vergelijking-A} \\
    & \dot{\vartheta}_\een + 2H\vartheta_\een=0. \label{eq:lege-vergelijking-B}
  \end{align}
\end{subequations}
Using~(\ref{FRW3}) and~(\ref{eq:exact-sol-mat}), the solution
of the system~(\ref{eq:lege-vergelijking}) is
\begin{subequations}
  \label{eq:lege-oplossing}
  \begin{align}
    \delta & =c_1t^{2/3}-2\psi t^{-1}, \label{eq:lege-oplossing-A}  \\
    \vartheta_\een & =-\dfrac{\nabla^2\psi}{a^2}. \label{eq:lege-oplossing-B}
  \end{align}
\end{subequations}
Since equations~(\ref{eq:lege-vergelijking-A})
and~(\ref{eq:lege-vergelijking-B}) are decoupled, the fluid flow has
no influence on the evolution of~$\delta$. Therefore,
$\vartheta^\true_\een=0$ so that $\vartheta_\een$ is equal to its
gauge mode.  This is to be expected, since in a pressure-less fluid
there is no fluid flow.  Notwithstanding the fact that there is no
local fluid flow, equation~(\ref{eq:lege-vergelijking-A}) yields a
growing mode.  This, however, is impossible.  As has been shown in
Section~$7^\ast$ \citep{miedema:2014ada}, $\vartheta_\een^\true=0$ and
$w=0$ yields the non-relativistic limit in which the gravitational
field is constant.  Consequently,
equation~(\ref{eq:lege-vergelijking-A}) is not correct.  It must be
concluded that the standard equation~(\ref{eq:lege-vergelijking-A})
cannot be considered as the non-relativistic limit of a cosmological
perturbation theory, as is assumed by
\cite{hwang-noh-1997,Hwang_2006}.  This fact will be further discussed
in the next subsection.

\subsection{Newtonian Theory of Gravity}
\label{app:newton}

It is generally assumed that if the energy density is dominated by
non-relativistic particles, so that $w\ll1$ and $\beta^2\ll1$, and if
the linear scales involved are small compared with the characteristic
scale $H^{-1}$ of the universe, then one may safely use the Newtonian
Theory of Gravity to study the evolution of density perturbations.

The perturbation equations of the (Newtonian) Jeans theory adapted to
an expanding universe after decoupling are given by \cite{c8},
equations (15.9.12)--(15.9.16). See also \cite{Mukhanov-2005},
Section~6.2 on the Jeans theory.  Substituting
$\boldsymbol{v}_1\coloneqq a\boldsymbol{u}_\een$,
$\rho\coloneqq\varepsilon_\nul$ and
$\rho_1\coloneqq\varepsilon_\een=\varepsilon_\nul\delta$ in
equations~(15.9.12)--(15.9.16) and taking the divergence of~(15.9.13),
one arrives at the Newtonian equations in the notation used in the
present article:
\begin{subequations}
  \label{subeq:mukhanov-textbook}
  \begin{align}
   & \dot{\delta}+\vartheta_\een =0, \label{muk1} \\
   & \dot{\vartheta}_\een+2H\vartheta_\een+\beta^2\dfrac{\nabla^2\delta}{a^2}+
          \dfrac{\nabla^2\phi}{a^2} =0, \label{muk2} \\
    &  \dfrac{\nabla^2\phi}{a^2} = \tfrac{1}{2}\kappa \varepsilon_\nul\delta, \label{muk3}
  \end{align}
\end{subequations}
where the energy conservation law~(\ref{FRW2}) with $w\ll1$ has been
used.  Differentiating (\ref{muk1}) with respect to time and
eliminating $\dot{\vartheta}_\een$ with the help of (\ref{muk2}) and,
subsequently, eliminating $\vartheta_\een$ and $\nabla^2\phi$ with the
help of (\ref{muk1}) and (\ref{muk3}), respectively, yields
\begin{equation}\label{eq:delta-standard-dust}
  \ddot{\delta} + 2H\dot{\delta}-
  \left[\beta^2\frac{\nabla^2}{a^2}+
   \tfrac{1}{2}\kappa\varepsilon_\nul\right] \delta =0,
\end{equation}
which is precisely the left-hand side of the standard
equation~(\ref{eq:delta-standard}) for $w\ll1$ and $\beta^2\ll1$.
Just as the system of
equations~(\ref{subeq:algemene-vergelijkingen-mat-dom}),
equation~(\ref{eq:delta-standard-dust}) is invariant under the
Newtonian gauge transformation~(\ref{eq:gauge-trans-newt}).  This
implies that the general solution of
equation~(\ref{eq:delta-standard-dust}) is gauge-dependent and the
gauge mode $\hat{\delta}$ given by~(\ref{eq:gauge-modes-standard}) is
for $w\ll1$ and $\psi=C$ one of the two linear independent solutions.
Consequently, the right-hand side of equation~(\ref{muk3}) is
gauge-dependent and $\phi$ depends on time.  As a consequence,
equation~(\ref{muk3}) is not the real Poisson equation.  It has been
shown in Section~7$^\ast$ \citep{miedema:2014ada} on the
non-relativistic limit that the source term of the Poisson equation is
invariant under the Newtonian gauge
transformation~(\ref{eq:gauge-trans-newt}), i.e., is gauge-invariant,
and the potential is independent of time.

The system of equations~(\ref{subeq:mukhanov-textbook}) will now be
compared with the relativistic system~(\ref{subeq:pertub-gi}) with
$w\ll1$ and $\beta^2\ll1$.  Using the expression~(45$^\ast$)
for~$R_\een$, the system~(\ref{subeq:pertub-gi}) becomes
\begin{subequations}
  \label{subeq:w-beta-klein}
  \begin{align}
    & \dot{\delta}+\tfrac{3}{2}H\delta+\vartheta_\een-
      \dfrac{4}{c^2}\dfrac{\nabla^2\phi}{a^2}\dfrac{1}{4H}=0,\label{eq:w-beta-klein-A}\\
    & \dot{\vartheta}_\een+2H\vartheta_\een+\beta^2\dfrac{\nabla^2\delta}{a^2}=0, \\
    & \dfrac{\nabla^2\dot{\phi}}{a^2}+2\kappa\varepsilon_\nul\vartheta_\een=0.
  \end{align}
\end{subequations}
Just as the system~(\ref{subeq:algemene-vergelijkingen}) follows
from~(\ref{subeq:pertub-gi}), the
system~(\ref{subeq:algemene-vergelijkingen-mat-dom}) follows
from~(\ref{subeq:w-beta-klein}).  Comparing the
systems~(\ref{subeq:mukhanov-textbook}) and~(\ref{subeq:w-beta-klein})
it must be concluded that the Newtonian
system~(\ref{subeq:mukhanov-textbook}) is not correct.

In the non-relativistic limit, i.e., $\vartheta_\een\rightarrow0$ and
vanishing pressure, the system~(\ref{subeq:w-beta-klein}) reduces~to
\begin{subequations}
  \label{subeq:w-beta-klein-non-rel}
  \begin{align}
    & \dot{\delta}+\tfrac{3}{2}H\delta-
     \dfrac{4}{c^2}\dfrac{\nabla^2\phi}{a^2}\dfrac{1}{4H} =0,\label{eq:w-beta-klein-non-rel-A}\\
    & \nabla^2\dot{\phi}=0.  \label{eq:phi-is-independent-of-time}
  \end{align}
\end{subequations}
Equation~(\ref{eq:phi-is-independent-of-time}) is equal to
equation~(53b$^\ast$) and implies that the Newtonian potential is
\emph{independent} of time, whereas
equation~(\ref{eq:w-beta-klein-non-rel-A}) is equal to
equation~(53c$^\ast$).  This can be shown as follows.  First,
eliminate $\nabla^2\phi$ using the expression~(45$^\ast$). Next, use
the perturbed Friedmann equation~(32a$^\ast$) with $\vartheta_\een=0$
to eliminate $R_\een$.  Finally, substitute
$\delta\coloneqq\varepsilon_\een/\varepsilon_\nul$ and use the
background equations~(\ref{subeq:einstein-flrw}).  The
non-relativistic limit is discussed in detail in Section~7$^\ast$
\citep{miedema:2014ada}.


\end{document}